# Interface engineering of van der Waals heterostructures towards energy-efficient quantum devices operating at high temperatures

Manh-Ha Doan[1*] and Peter Bøggild[1*]

[1] Department of Physics, Technical University of Denmark, Kgs. Lyngby 2800, Denmark

*E-mail: mando@dtu.dk and pbog@dtu.dk

**Abstract**

Quantum devices, which rely on quantum mechanical effects for their operation, may offer advantages, such as reduced dimensions, increased speed, and energy efficiency, compared to conventional devices. However, quantum phenomena are typically observed only at cryogenic temperatures, which limits their practical applications. Two-dimensional materials and their van der Waals (vdW) heterostructures provide a promising platform for high-temperature quantum devices owing to their strong Coulomb interactions and/or spin-orbit coupling. In this review, we summarise recent research on emergent quantum phenomena in vdW heterostructures based on interlayer tunnelling and the coupling of charged particles and spins, including negative differential resistance, Josephson tunnelling, exciton condensation, and topological superconductivity. These are the underlying mechanisms of energy-efficient devices, including tunnel field-effect transistors, topological/superconducting transistors, and quantum computers. The natural homojunction within vdW layered materials offers clean interfaces and perfectly aligned structures for enhanced interlayer coupling. Twisted bilayers with small angles may also give rise to novel quantum effects. In addition, we highlight several proposed structures for achieving high-temperature Majorana zero modes, which are critical elements of topological quantum computing. This review is helpful for researchers working on interface engineering of vdW heterostructures towards energy-efficient quantum devices operating above liquid nitrogen temperature.

# 1. Introduction

Energy consumption is a crucial issue in modern electronics, specifically with high demands for applications in emerging fields, such as artificial intelligence, high-performance computing, and internet communication [1-4]. Field-effect transistors (FET), the fundamental building blocks of computer chips, have been scaled down exponentially in recent decades, a trend known as Moore's law [5, 6]. This scaling, together with the innovation of radical device designs and integration techniques, for example Fin-FET, gate-all-around FET, and monolithic 3D integration, has boosted computing speed while maintaining constant power consumption [7-9]. However, as the operation of conventional transistors is based on the modulation of the current flow in the device by an electrical field, there are critical challenges in maintaining good performance, reliability, and manufacturability when the device dimensions are reduced to the few-nanometre scale [10-12]. Surface states, heat dissipation, short-channel effects, and quantum confinement effects are some of the bottlenecks of conventional FET technology based on silicon and III-V semiconductors [13-17]. Searching for new materials and alternative computing paradigms, therefore, would be a solution to overcome these challenges for future electronics.

In a conventional metal–oxide–semiconductor FET, the current flow is switched on and off through the thermionic injection of electrons over a gate-tunable energy barrier. A gate voltage of at least 60 mV is required to increase the current by one order of magnitude at room temperature [18]. This value is typically termed Boltzmann's tyranny [19].

Quantum devices, including tunnel FET (TFET) [18, 20], topological transistors [19, 21], and superconducting transistors [22, 23], are predicted to overcome Boltzmann's tyranny. In TFETs, current flows *via* quantum tunnelling through a barrier rather than thermionic emission, allowing the achievement of a subthreshold slope below 60 mV/decade, which in principle can be as low as zero [18, 20, 24]. In addition, TFETs have much lower leakage currents because of their steep subthreshold slope and the fact that the tunnelling mechanism is more controllable at low voltages, resulting in remarkable power savings. Topological/superconducting transistors rely on electrically controlled quantum phase transitions in topological insulators or superconductors. Because spin and charge transport can occur without dissipation in the conducting channel, the device performance is limited only by the power dissipated when switching the transistor on and off. This process is theoretically predicted to be energy efficient compared to that of a conventional MOSFET [19, 25-27].

In a classical computer, FETs act as building blocks with ON and OFF states representing binary information (1s and 0s), allowing the execution of logical operations necessary to perform calculations. Quantum computing is a cutting-edge field that leverages the principles of quantum mechanics to process information in fundamentally different ways [28-30]. The basic unit of quantum information is a qubit, which can represent both 0 and 1 at the same time due to superposition. This allows different computational tasks to be performed simultaneously, which, for some computation tasks, leads to an exponential speed increase. Quantum computers are expected to produce solutions to complex problems which cannot be solved by classical computers or that demand extreme amounts of computational and energy resources. Among numerous qubit implementations, topological qubits are predicted to be more stable and inherently robust compared to popular approaches based on elementary particles such as trapped ions and single photons/electrons/spins [31-34]. Topological qubits are thus promising candidates for building large-scale quantum computers that are less susceptible to noise errors and therefore have higher error resistance. Experimentally, topological qubits can be realised in topological superconducting structures by placing a conventional superconductor in proximity to a material with strong spin-orbit coupling, such as topological insulators, semiconductor nanowires, or chains of magnetic atoms chains [35-37].

Tremendous efforts and resources from both academia and mainstream technology companies are spent on searching for innovation for post-Moore era electronics [38, 39]. TFETs, topological insulators, and topological qubits have been realised in epitaxially grown heterostructures made of conventional semiconductors and superconductors, such as silicon and III-V compounds [40-42]. However, fabricating high-quality nanostructures without defects and dislocations at the surfaces and interfaces of these materials *via* epitaxial growth is a great challenge. Even for near-perfect heterostructure interfaces, quantum effects are rarely observed at high temperatures (above 77 K), which severely limits their practical applications [43, 44].

Two-dimensional (2D) layered materials and their heterostructures, the so-called van der Waals (vdW) heterostructures, have drawn extensive attention recently owing to the intriguing possibility of engineering heterostructures with atomic precision without concern for lattice mismatch and interface disorder, as well as their possible use in future electronic components [45-47], circuits [48], and architectures [49]. 2D materials are atomically thin and offer a smooth channel-to-dielectric interface without dangling bonds. Relatively high carrier mobilities combined with low leakage currents can be achieved even for channel thicknesses below one nanometer, which are unable to achieve in Si-based devices [49, 50]. Moreover, 2D materials can be stacked on arbitrary substrate or made freestanding. This offers the possibility of monolithic 3D integration [51]. Importantly, in 2D materials and vdW heterostructures, sharp interfaces, strong Coulomb interactions, and/or spin-orbit coupling permit observations of various quantum phenomena even at room temperature [52-56].

In this article, we review recent progress in the interface engineering of vdW heterostructures to realise high-temperature quantum effects based on interlayer tunnelling and the coupling of charged carriers and spins for energy-efficient devices, including TFETs, topological/superconducting transistors, and quantum computers. Interface engineering is defined as i) selecting 2D components appropriate for the target heterostructure and of sufficient quality, ii) stacking them at right angles, and iii) maintaining clean and sharp interfaces to maximise interlayer coupling and reduce detrimental scattering effects. The studied emergent phenomena include quantum tunneling, Josephson tunneling, exciton condensation, and topological superconductivity, as depicted in Fig. 1. In section 2, we discuss the phenomenon of negative differential resistance, which originates from the band-to-band tunnelling of single electrons/holes in vdW heterostructures. This is the mechanism behind the operating principle of the TFET. Quantum tunnelling through a thin barrier also occurs with paired particles, as in the case of superconducting junctions, known as Josephson tunnelling. Josephson junctions (JJs) have important applications for superconducting quantum interference devices (SQUIDs), superconducting transistors, and quantum circuits [57-61]. JJs built from vdW heterostructures are addressed in section 3. Under certain conditions, the bound pairs of electrons and holes between two separated 2D layers form a macroscopic quantum state of Bose-Einstein condensation (BEC), akin to Cooper pairs in superconductors [62, 63]. The condensation of excitons has been explored as a potential avenue for achieving high-temperature superfluidity and superconductivity [63-66]. We present the recent processes in searching for room-temperature exciton BEC in vdW heterostructures in section 4. In section 5 we review high-temperature quantum spin Hall effects in 2D materials and their potential applications for topological transistors and spin FETs. Topological superconductivity in 2D materials and vdW heterostructures, which may host Majorana modes for topological quantum computing is discussed in section 6. Several theoretically proposed structures for seeking high-temperature Majorana zero modes in vdW heterostructures are also addressed in this section. Finally, in the last section, we summarise current state of the art and offer perspectives for the research field looking forward.

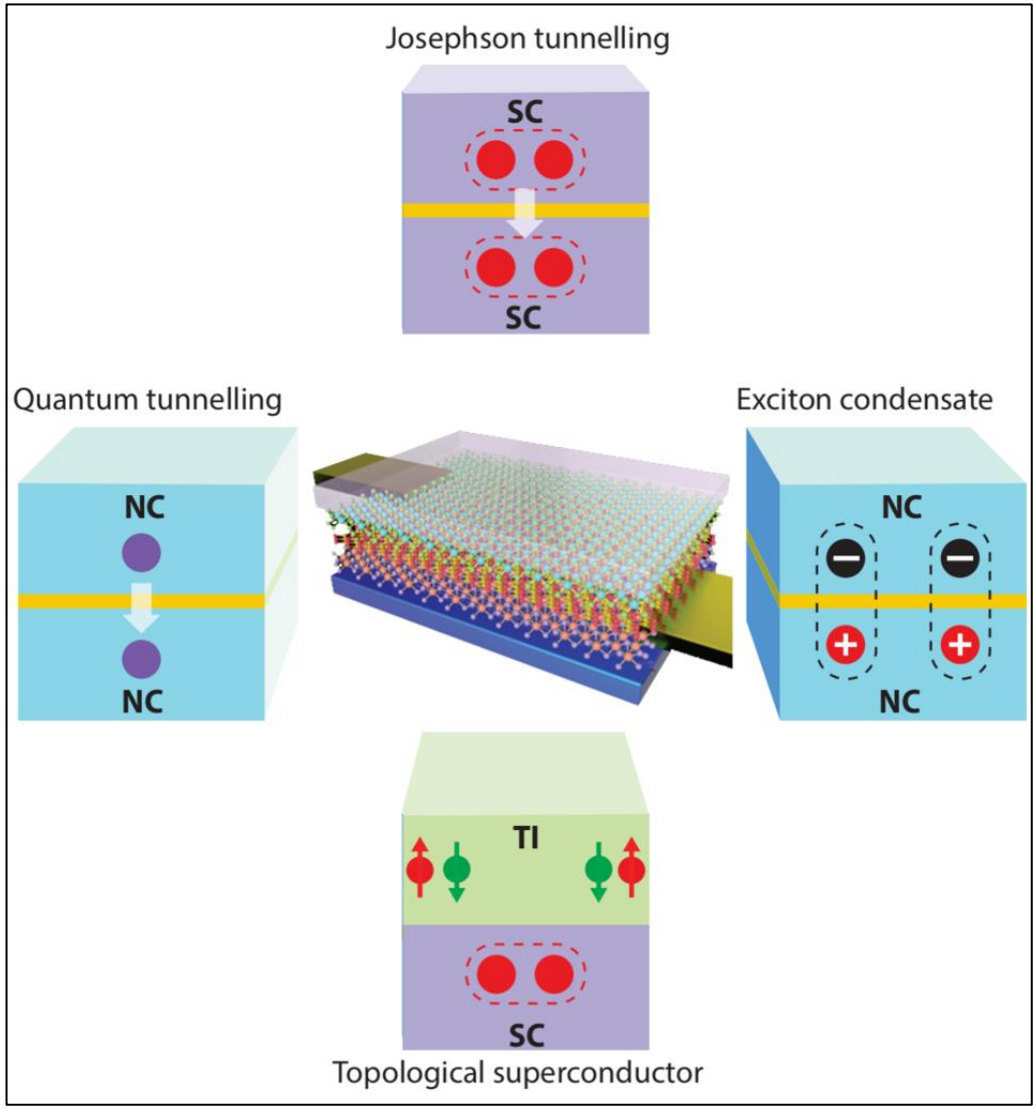

**Figure 1. Emergent phenomena in vdW heterostructures due to interlayer tunnelling and the coupling of charged carriers and spins including quantum tunnelling, Josephson tunnelling, exciton condensate, and topological superconductivity.** These are the underlying mechanisms for energy-efficient devices such as tunnel field-effect transistors, topological/superconducting transistors, high-temperature superconducting devices, and quantum computers. The center panel illustrates a typical device with two conducting monolayers separated by a few layers of an insulator. Two gold electrodes are used to inject charge carriers into the top and bottom conducting layers. NC: normal conductor (semiconductor, semimetal), SC: superconductor, TI: topological insulator.

## 2. Negative differential resistance and tunnel field-effect transistor

The current-switching process of an FET is characterized by a parameter called the subthreshold swing (SS), which is defined as $SS = \frac{\partial V_G}{\partial \Psi} \frac{\partial \Psi}{\partial (log_{10} I_D)}$, where $V_G$ is the gate voltage, $I_D$ is the drain current, and $\Psi$ is the surface potential of the semiconductor [18]. The first term is the transistor body factor and equals the unit in the ideal case, whereas the second term reflects the conduction mechanism in the channel. For conventional FETs based on thermionic injection, the minimum value of the second factor can be calculated as $\frac{kT}{q} ln10$, where k, T, and q are the Boltzmann constant, temperature, electron charge, respectively, which has a value of 60 mV/dec at room temperature [18]. TFETs, where the current flow is governed by quantum-mechanical band-to-band tunnelling, would achieve an SS value lower than 60 meV (ref. 18, 20). Experimentally, SS can be extracted from the slope of the transfer curve ($I_D$ vs. $V_G$) in the subthreshold region [18]. TFETs offer not only a higher on/off ratio but also power savings with a smaller SS compared to those of other devices, including bulk silicon-based metal-oxide-semiconductor FETs and high-mobility FETs [18].

The heart of TFETs is an Esaki diode, where band-to-band tunnelling is manifested by a negative differential resistance (NDR) in the current-voltage curves of the device owing to the alignment of the energy bands of the junction [67]. Esaki diodes can be built from a heavily doped p-n junction, forming a narrow depletion region or broken-gap heterostructure with ultrathin tunnel barriers [67-69]. In the tunnelling regime, the current reaches a maximum value when the band edges of the junction are well aligned, and then decreases with increasing applied voltage, resulting in NDR behaviour.

Owing to the excellent band-edge modulation via gating in 2D materials, TFETs have been realised in numerous vdW heterostructures [70-78]. For example, by stacking bilayer $MoS_2$ on highly p-doped Ge, Sarkar *et al.* reported a TFET with a subthermionic subthreshold swing at room temperature [70]. Because the band alignments of the Ge-$MoS_2$ junction form a staggered heterostructure, the applied gate voltages modulate the band edges of $MoS_2$, which in turn, controls the tunnelling current through the junction. Although the transfer curves of the fabricated devices show sub-60 mV SS, no clear NDR behaviour was observed in their I-V characteristics [70]. This is just an unusual slop region in the I-V curves, which the authors addressed as a trend toward

NDR, as MoS$_2$ is only slightly n-doped [70]. However, this may be because of the numerous defects formed in Ge/MoS$_2$ during the stacking process.

Interface quality is a critical factor in tunnel junctions. Defects at the interface can act as trap and/or recombination centres, leading to reduced tunnelling current [79, 80]. They may also induce pinning effects between band edges, resulting in junction blurring [81]. In vdW heterostructures fabricated by manual stacking processes, defects are commonly created at the interface during sampling in ambient conditions, regardless of using the so-called wet or dry transfer methods [82, 83]. These interfacial defects include bubbles of air, water vapour, and/or hydrocarbon residues trapped at the interface [84-86]. Creating a clean vdW heterostructure interface would be a route to achieving high-performance TFETs. Indeed, the junctions of WSe$_2$/MoSe$_2$ and MoS$_2$/WSe$_2$ synthesized by metal-organic chemical vapor deposition (MOCVD) exhibited robust NDR behavior at room temperature [75], as shown in Fig. 2a and 2b. While some progresses have been made in high-quality 2D crystals [87], the synthesis of large-scale vdW heterostructures with defect-free interfaces for device applications remains highly challenging [88-90]. By selecting 2D components to form a largely staggered junction, clear NDR characteristics can be observed in stacked vdW heterostructures [76]. For example, Fig. 2c and 2d show the structure and I-V curves of the SnSe$_2$/black phosphorous (BP) heterostructure [76]. Using high-quality thin insulating layers, such as h-BN or Al$_2$O$_3$, inserted between the two 2D semiconducting layers may protect the junction sharpness, thereby enhancing the tunnelling current [77, 78] (Fig. 2e and 2f). Fabricating vdW in an inert environment also improves the NDR behaviour of the device [91, 92].

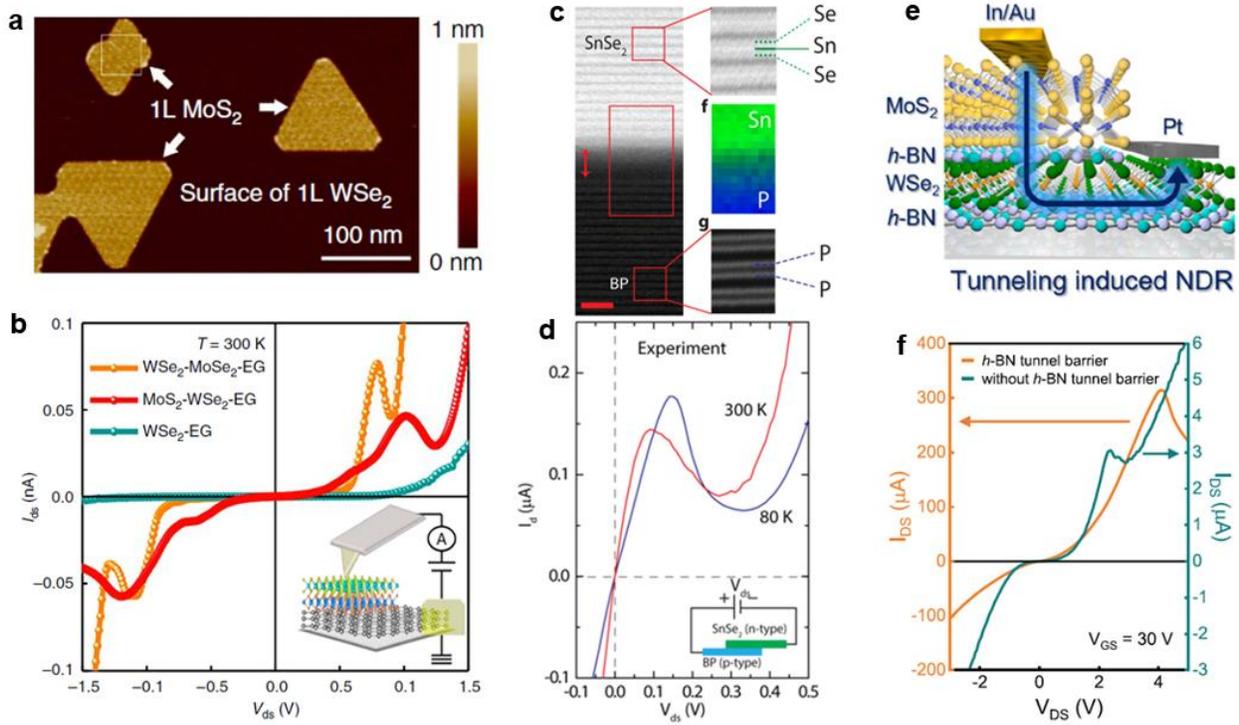

**Figure 2. Observation of negative differential resistance (NDR) at room temperature in vdW heterostructures.** (a, b) AFM image and I-V characteristics of synthetic $MoS_2$/$WSe_2$ vdW heterostructures grown by metal-organic chemical vapour deposition (MOCVD). Synthetic vdW heterostructures offer clean interfaces, but are typically limited to small scales and few selected 2D components. EG: multilayer epitaxial graphene. (c, d) Cross-sectional TEM images and I-V characteristics at 80 and 300 K of stacked $SnSe_2$/BP. BP: black phosphorus. A thin amorphous region formed at the interface during device processing acts a tunnelling barrier and maintains the broken-gap band alignment of the heterostructure for observation of the prominent NDR. Scale bar in (c) is 2 nm. (e, f) Device schematic and I-V characteristics of the stacked $MoS_2$/hBN/$WSe_2$ heterostructures at room temperature. Insertion of thin insulating hBN layers (thickness of 3-4 nm) at the interface results in the formation of sharp band edges and an inelastic tunnelling current. NDR with a typically high peak current (315 µA) is achieved at room temperature. Reprinted and adapted with permission: (a) and (b) from [75] (Springer Nature), (c) and (d) from [76] (ACS), (e) and (f) from [77] (ACS).

Van der Waals heterostructures made of ultrathin crystals from 2D materials can be used to build resonant tunneling devices, which have potential applications for the single-electron FET [93-99]. Graphene/hBN/graphene junctions are among the most frequently investigated vdW heterostructures for resonant tunneling and NDR devices, because they offer inert and clean interfaces [93, 94]. However, graphene is a gapless semiconductor, which limits the tunability of its gate modulation in TFETs. Building the channel part with 2D few-layer semiconductors, semiconducting vdW heterostructures, or moiré superlattices allows the generation of multiple energy levels for resonant tunnelling *via* layer number control, gate modulation, and twist angle engineering [95-98]. Fig. 3 shows some examples of TFET with multiple NDR peaks based on few-layer crystals of InSe and WSe$_2$ [97, 98]. Observations of room-temperature resonant tunnelling in these vdW heterostructures provide an opportunity to build single-electron transistors for practical applications [99, 100].

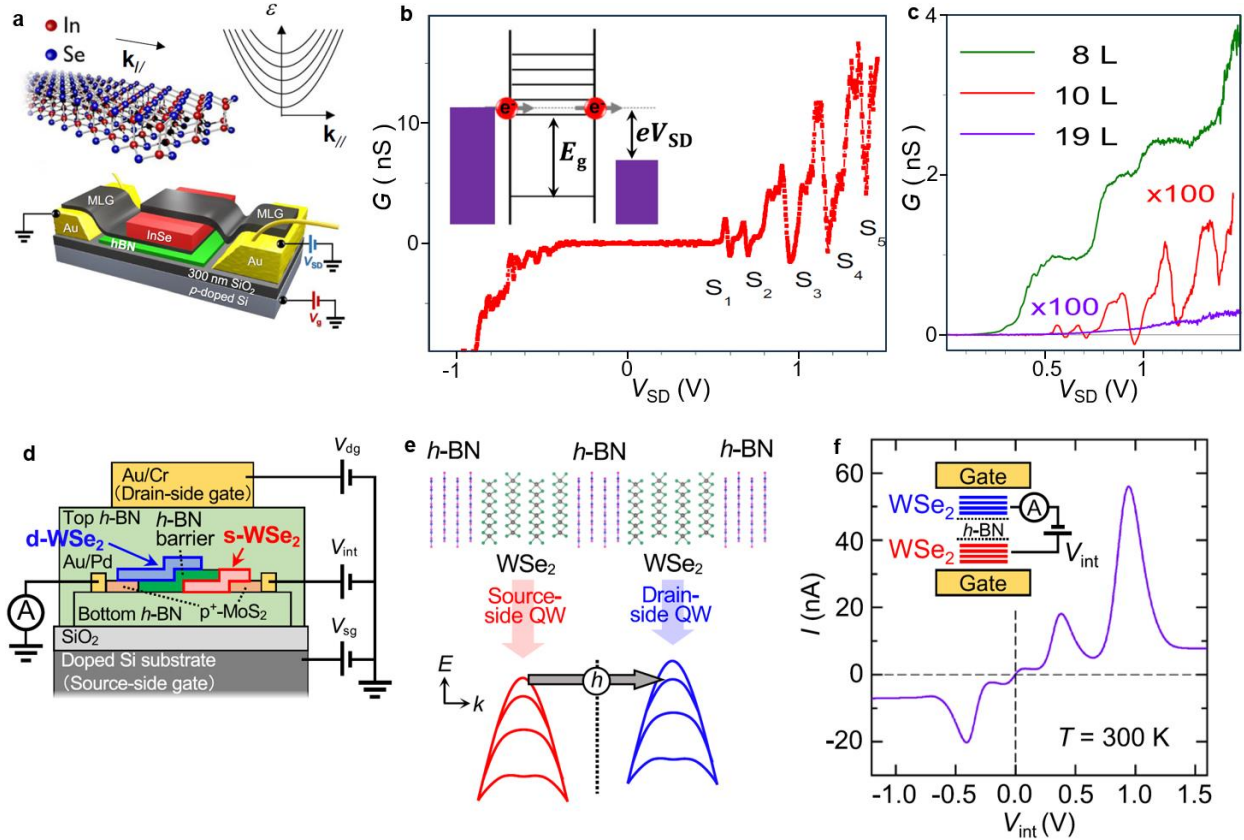

**Figure 3. Observation of resonant tunnelling into subbands of multilayer vdW materials.** (a) Crystal structure and device schematic for observation of resonant tunnelling in multilayer InSe. MLG: multilayer graphene. The energy of the 2D subbands can be tuned by varying the thickness of the InSe layer, owing to strong quantum confinement of the light-effective mass electrons. Resonant tunnelling occurs when the chemical potential in the source (MLG electrodes) is matched with a given 2D subband of InSe, leading to a peak in the I-V characteristics of the device. (b) Differential conductance curve, G vs. $V_{SD}$, at $V_g = 0$ V and T = 2 K for a device with ten-layer InSe. The inset shows the tunnelling of electrons from the MLG electrode (source) into the 2D subbands of InSe. (c) Differential conductance curves of devices with InSe thicknesses of 8, 10, and 19 layers at $V_g = 0$ V and T = 2 K. (d-f) Device schematic, band structures, and I-V curve at 300 K of the WSe$_2$/hBN/WSe$_2$ vdW heterostructure. The few-layer p-doped WSe$_2$ (3-5 layers) source and drain act as the quantum well with quantised subbands for holes, while a thin hBN (3-6 layers) is inserted at the interface to be used as the tunnelling barrier. Multiple NDR peaks are observed in I-V curve at 300 K and $V_{backgate} = -70$ V and $V_{topgate} = -13.5$ V. Reprinted and adapted with permission, (a-c) from [97] (Springer Nature), (d-f) from [98] (ACS).

One of the great advantages of vdW heterostructures over conventional epitaxial heterostructures is that there are no dangling bonds at the interface. This offers the ability to control the twist angle between the layers during device fabrication, which strongly alters the interlayer tunneling [101-103]. Fig. 4 shows that the tunnelling current in a graphene/hBN/graphene structure changes significantly when the twist angle between the two graphene layers is adjusted even a fraction of a degree [102]. The NDR peaks vanish when the twist angle is larger than $5^o$. This is attributed to alignment of the Dirac cones between the top and bottom graphene layers [102]. In vdW heterostructures made of 2D semiconductors or semimetals with anisotropic conductivity, tunneling currents should also depend strongly on the twist angle due to the alignment/misalignment of the Fermi surfaces or Fermi arcs [103, 104]. For example, Srivastava *et al.* observed that the transport properties of stacked black phosphorus junctions transform from linear to tunnelling behavior when the twist angle is changed from $0^o$ to $90^o$ [ref. 103]. Therefore, attention should be paid to the twist angle during device processing to optimise the tunnelling current and NDR peaks.

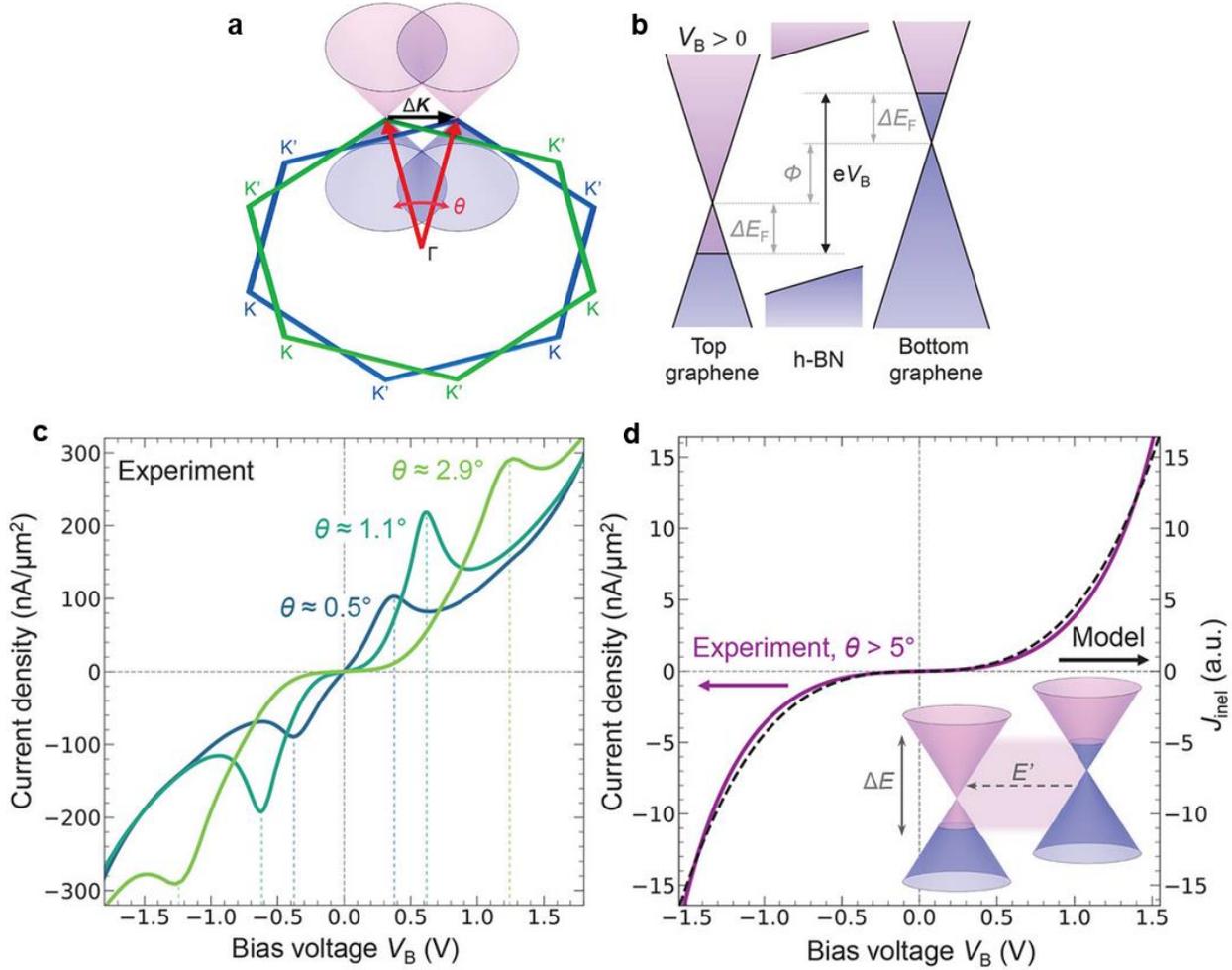

**Figure 4. Twist-angle dependence of resonant tunnelling in vdW heterostructures**. (a) Alignment of Brillouin zones of two graphene layers stacked at a twist angle of $\theta$. The twist angle translates into a wavevector mismatch $\Delta K$ between the Dirac points of the top (green) and bottom (blue) graphene monolayers. (b) Energy band alignment of the Gr/hBN/Gr heterostructure under an applied bias between the top and bottom graphene layers. The applied voltage $V_B$ introduces an energy offset $\Phi$, together with the Fermi level shift $\Delta E_F$. An increase in the twist angle enlarges the separation of the Dirac cones in momentum space and shifts the resonance condition to higher $V_B$ values. For $\theta \gtrsim 5°$, the momentum mismatch $\Delta K$ becomes large, leading to vanishing tunnelling current peaks. (c, d) I-V characteristics of graphene/hBN/graphene tunneling devices with different twist angles. Reprinted and adapted with permission, (a-d) from [102] (ACS).

As the tunnelling current is strongly affected by defects at the interface and twist angles between the layers, a natural homojunction formed within vdW materials can offer a clean and well-aligned interface for achieving high-performance TFETs [105, 106]. Black phosphorous (BP) is a good candidate for this purpose because it is a semiconductor with high carrier mobility, a direct band gap even in the bulk form, and strongly thickness-dependent band edges [107]. Indeed, by making the device within the same flake of BP with different thicknesses between the monolayer and multilayer regions, Kim *et al.* built a TFET with superior performance compared to those of the previous ones [108], as summarised in Fig. 5. Their TFETs show record-low average SS values (~ 22.9 mV dec$^{-1}$) over 4 current decades with record-high on-current ($I_{60}$ = 0.65 µA µm$^{-1}$), which are promising for application in low-power switching devices [108]. However, BP is very sensitive to the oxygen environment; thus, passivation methods and/or special device processing should be developed to prevent device degradation for real applications [109-111].

Alternative methods have also been developed in recent years to achieve subthermionic-SS FETs such as negative-capacitance FETs [112-114], phase-transition FETs [115, 116], Dirac-source FETs [117-119], and junctionless NDR devices [120, 121]. However, while negative-capacitance and phase-transition FETs usually exhibit hysteresis in the switching process, others may require complicated device configurations with additional fabrication steps.

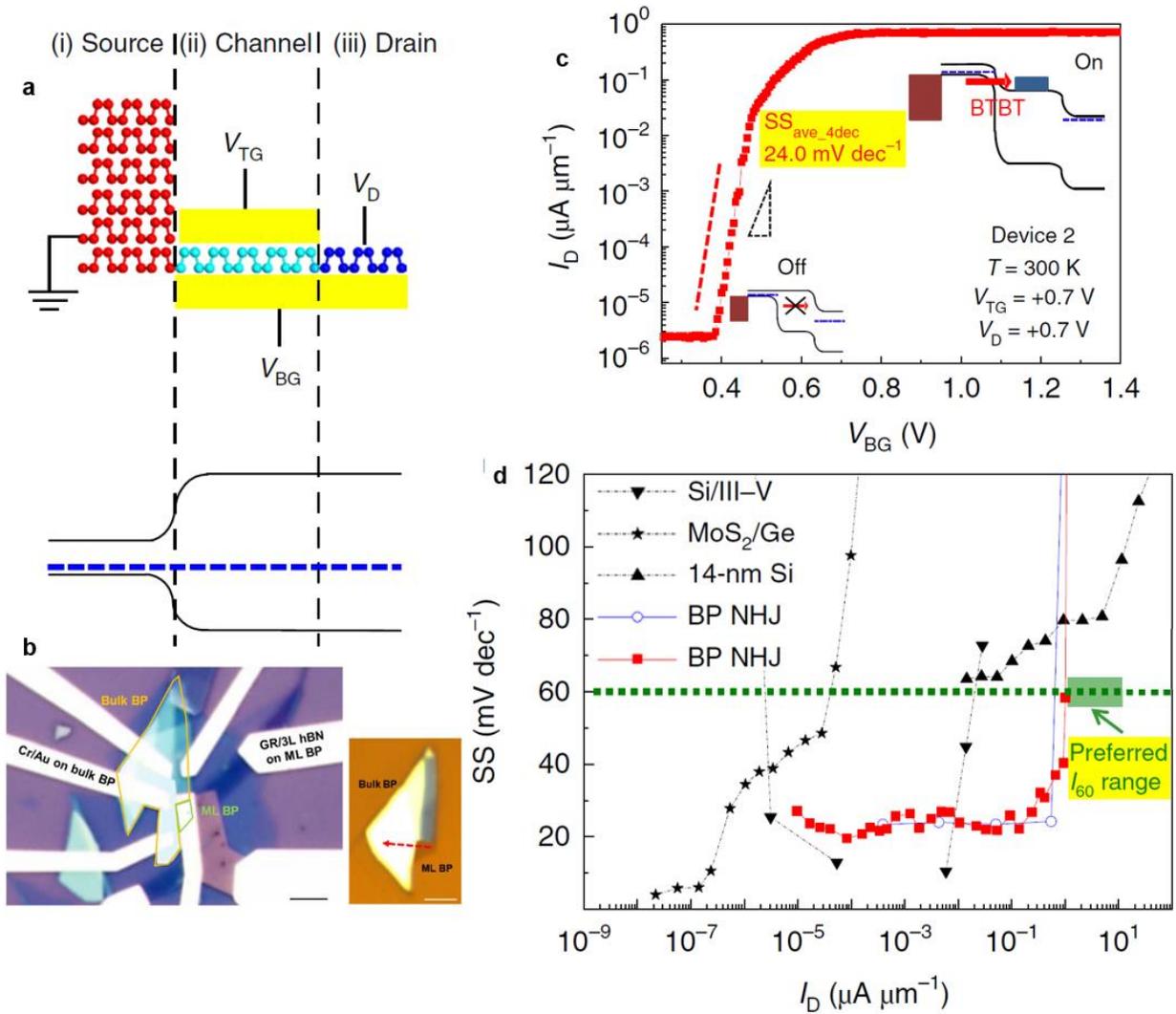

**Figure 5. Natural homojunction TFET.** (a) Schematic structure and band diagram of a TFET made of black phosphorus (BP) with different thickness regions between the bulk and monolayer within the same exfoliated flake. (b) Optical image of the BP tunnelling device. Graphene (GR) and 3-layer hBN are used as the back-gate and dielectric layer for monolayer BP (ML BP) region, respectively. c) Transfer characteristics ($I_D$ vs. $V_G$) of device with bottom gate control. (d) Comparison of the subthreshold swing (SS) as a function of the ON current of the BP TFET with that of various FETs showing the achievement of sub-60 meV SS with a high ON current. Reprinted and adapted with permission, (a-d) from [108] (Springer Nature).

## 3. Josephson tunneling junctions and diodes

In superconductors electrons form Cooper pairs, that maintain quantum coherence over macroscopic distances and move through a material without resistance [122, 123]. When two superconductors are separated by a thin insulating layer, the Cooper pairs can tunnel through this barrier allowing a supercurrent to flow between the superconductors without any applied voltage, an effect called Josephson tunneling as it predicted in 1962 by Brian D. Josephson [124].

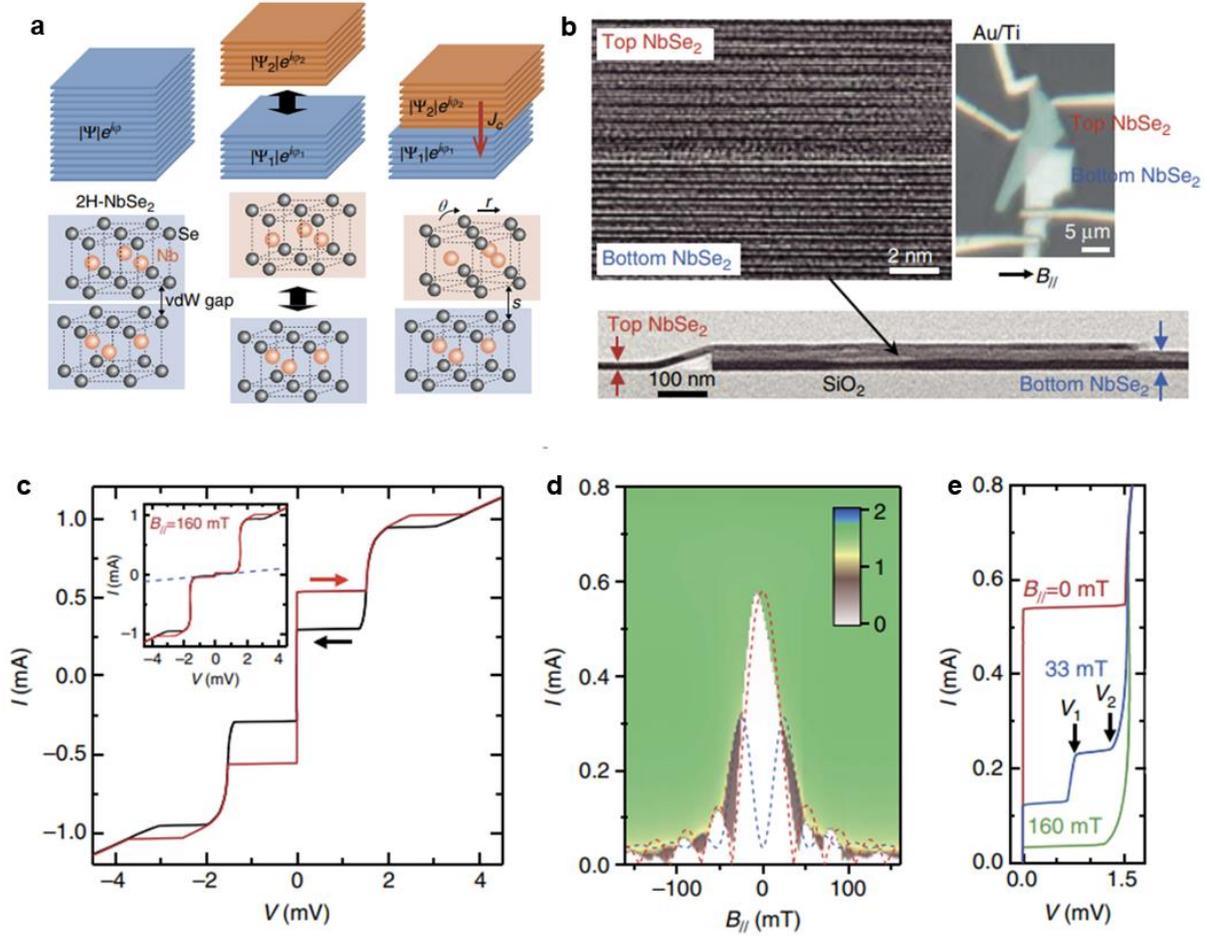

**Figure 6. vdW Josephson junction.** (a, b) Crystal structure and cross-sectional TEM images of the $NbSe_2/NbSe_2$ junction. In the natural bulk form, the interconnected layers function as one superconductor with a single-order parameter $|\Psi|e^{i\varphi}$. In stacked junctions, misalignment and misorientation of the layers and/or a larger vdW gap in the stacked junction induce interlayer decoupling and require different order parameters for each side of the vdW gap, $|\Psi_1|e^{i\varphi_1}$ and $|\Psi_2|e^{i\varphi_2}$. (c) Electrical characteristics of the device showing the characteristics of the JJ with a large critical current of $I_c = 0.53$ mA and hysteresis in the I–V curve. (d) Contour plot of V as a function of I and $B_{//}$ at 2 K, showing the Fraunhofer pattern of JJ. The colour bar shows the value of V in mV. (e) I-V curves of the junction with applied in-plane B. Fiske resonances are observed as current steps at bias voltages $V_1$ and $V_2$ when $B_{//} = 33$ mT. Reprinted and adapted with permission, (a-e) from [125] (Springer Nature).

A simple vdW JJ can be fabricated by stacking 2D superconductors, for example $NbSe_2$, together even without an inserted insulating layer [125], as shown in Fig. 6. In the natural bulk form, the $NbSe_2$ multilayers have the same atomic arrangement, and the interconnected layers work as one superconductor with a single-order parameter, $|\Psi|e^{i\varphi}$, as shown in Fig. 6a. Manually stacking the two layers together induces misalignment, misorientation, and/or a larger vdW gap (Fig. 6a). When the interlayer decoupling becomes sufficiently large, the superconducting state of the $NbSe_2$ crystal created across the artificially connected vdW interface cannot be described by a single-order parameter but rather requires different order parameters for each side of the vdW gap, $|\Psi_1|e^{i\varphi_1}$ and $|\Psi_2|e^{i\varphi_2}$. The flow of the supercurrent across the interface then follows the Josephson relation: $I = I_c\sin(\varphi_1-\varphi_2)$, where $I_c$ is the critical current [125]. Although there is a blurred region observed at the interface of the junction (Fig. 6b), possibly attributed to the stacking process being carried out under ambient conditions, the fabricated $NbSe_2/NbSe_2$ device shows a large critical current of $I_c = 0.53$ mA with hysteresis of the I–V curve, characteristics of the JJ (Fig. 6c). In addition, the Fraunhofer pattern and Fiske resonance are also observed in the magnetic field dependence of the I–V curve of the device (Fig. 6d and 6e), further confirming the properties of the JJ [125].

The atomic and tunable thickness of 2D materials with a broad conductivity range, from insulating to metallic and superconducting [126-128], allows us to choose various types of barriers for JJ to explore new tunnelling mechanisms and novel quantum devices [129-134, 137, 138]. For example, by inserting a monolayer ferromagnet, $Cr_2Ge_2Te_6$, between superconducting $NbSe_2$, Josephson tunnelling becomes spin-dependent and can be modulated by the magnetization direction of the barrier [132], as shown in Fig. 7a and 7 b. In $NbSe_2$, two spin components of Cooper pairs remain localized predominantly inside each layer, (K↑ - K' ↓ or K↓ - K' ↑, where K and K' denote the electronic band near the K and K' points) due to Ising Cooper pairing [132]. With perpendicular magnetization, Ising-Cooper pairs can tunnel via spin-dependent energy levels without a spin flip, whereas for in-plane magnetization, tunnelling occurs with a spin flip. These results may lead to new spin- and phase-controlled quantum electronic devices [135, 136]. Using three-layer $Nb_3Br_8$ as the tunnel barrier, field-free superconducting diode behavior, where a supercurrent and a normal current flow in the opposite directions in the absence of an applied magnetic field, is observed in the $NbSe_2/Nb_3Br_8/NbSe_2$ heterostructures [137], as shown in Fig. 7(c-e). Because the unit cell of $Nb_3Br_3$ is composed of six layers of Nb-Br edge-sharing octahedra with inversion centres lying between adjacent layers, the inversion symmetry is preserved in the even-layer crystal but broken in the odd-layer structure [137]. The asymmetric I-V characteristic (non-reciprocal behaviour) of JJ is due to the breaking of inversion symmetry in the vdW heterostructure, which consists of three layers of $Nb_3Br_3$ [137].

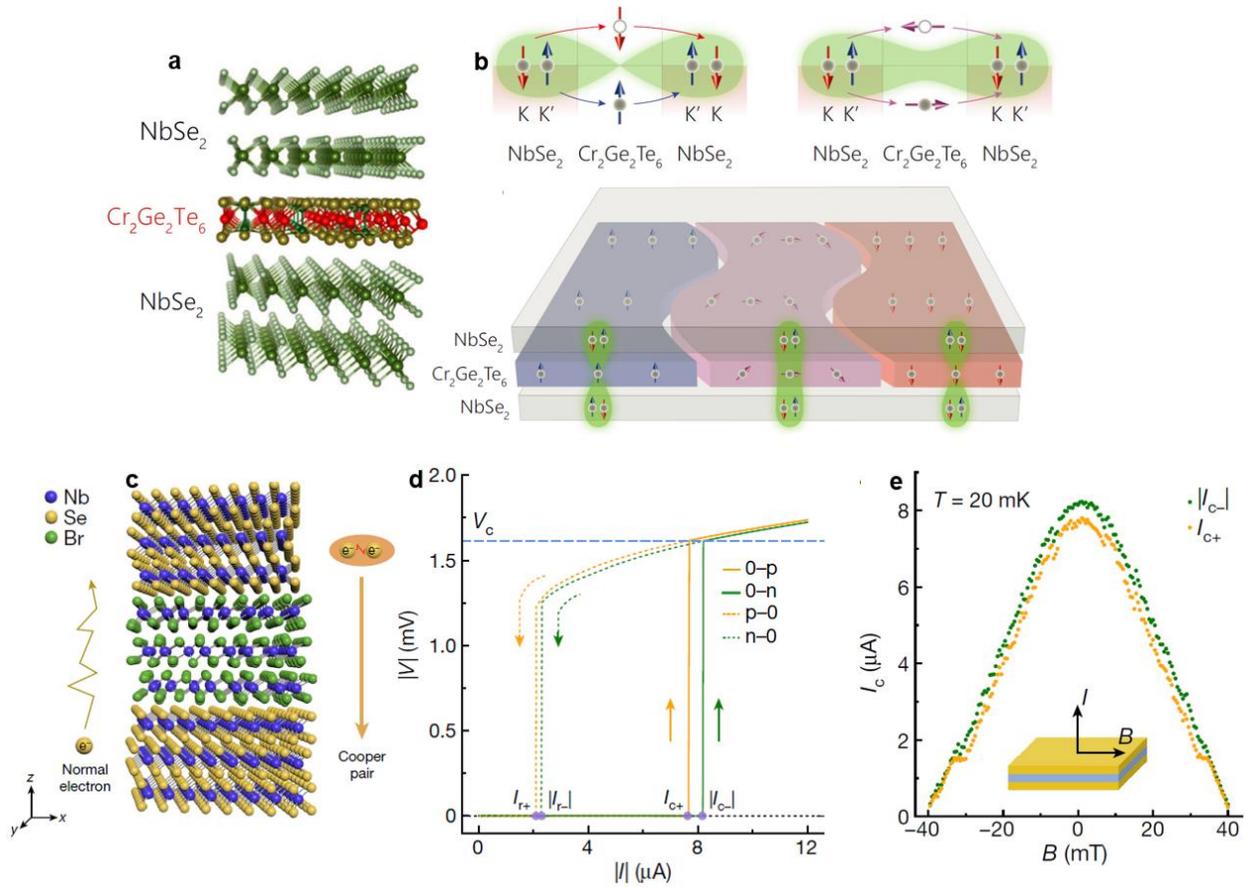

**Figure 7. Magnetic JJ and field-free vdW Josephson diodes.** (a, b) Crystal structure and spin-dependent tunnelling processes in the $NbSe_2/Cr_2Ge_2Te_6/NbSe_2$ JJ. $NbSe_2$ is a superconductor, while $Cr_2Ge_2Te_6$ is a magnetic insulator. Under perpendicular magnetisation, Ising-Cooper pairs can tunnel via spin-dependent energy levels without a spin flip, whereas for in-plane magnetisation, tunnelling occurs with a spin flip. (c-e) Crystal structure and electrical characteristics of the field-free Josephson diode made of $NbSe_2/Nb_3Br_8/NbSe_2$ heterostructures. The supercurrent flows in one direction, whereas the normal current flows in the other direction because of the broken inversion symmetry of the vdW heterostructure owing to the odd layers of $Nb_3Br_8$. The measured I-V curve in (d), which contains four branches: sweeping the current from zero to positive (0–p), from positive back to zero (p–0), from zero to negative (0–n), and from negative back to zero (n–0), shows a large hysteresis with similar critical currents for the negative and positive regimes ($I_{c+}$ vs. $|I_{c-}|$ or $I_{r+}$ vs. In panel (e) $|I_{r-}|$. $I_{c+}$ and $|I_{c-}|$ are obtained from the 0–p and 0–n branches, respectively, while the magnetic field is swept from positive to negative in (e). The diode effect 'turns off' by ±35 mT. Reprinted and adapted with permission, (a) and (b) from [132] (Springer Nature), (c-e) from [137] (Springer Nature).

The discovery of high-temperature superconductivity in monolayer $Bi_2Sr_2CaCu_2O_{8+\delta}$ (BSCCO) provides an ideal 2D quantum system for understanding unconventional superconductivity as well as for potential quantum devices operating above liquid nitrogen temperatures [139]. BSCCO is a member of the cuprate family and has been known for its high-temperature superconductivity since the 1980s. However, producing high-quality thin layers from its bulk form is extremely challenging owing to the sensitivity of their surfaces to the air environment. By developing sample fabrication processes where BSCCO flakes are exfoliated on a cold stage kept at -40 ºC inside an Ar-filled glove box with water and oxygen content below 0.1 ppm, monolayer superconducting with Tc as high as that of the bulk form (88 – 90 K) is achieved, as shown in Fig. 8a. Importantly, the superconductivity of monolayer can be highly tunable by doping *via* annealing processes to modulate its oxygen concentration (Fig. 8b-c). This opens a great opportunity for the fabrication of electrically controlled superconducting devices such as Josephson diodes and transistors, quantum magnetometry, and quantum circuits [140-145]. Indeed, there have been recent studies reported about high-temperature Josephson diodes made of BSCCO stacked homojunctions [146] as well as gate-modulated superconductivity in thin BSCCO layers [147].

Importantly, Josephson tunnelling and diode behaviour are found to be sensitive to the stacking angles between the BSCCO layers [146], suggesting that twist angle engineering is an interesting and promising route for investigating novel quantum phenomena in these systems at high temperatures for practical applications, such as interferometers and quantum qubits [148, 149].

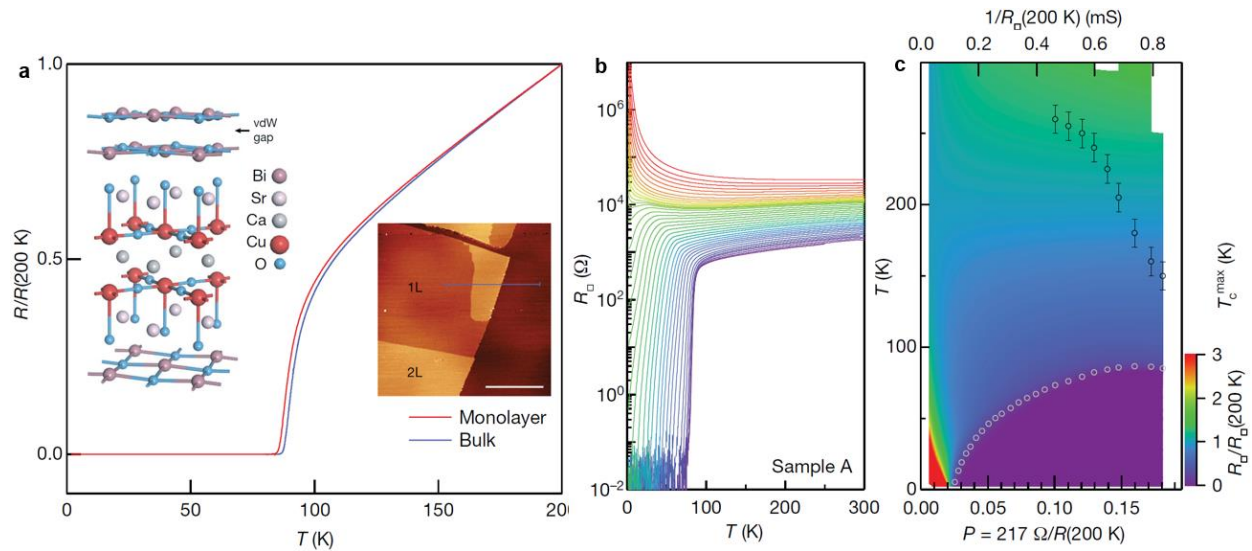

**Figure 8. Tunable high-temperature 2D superconductivity.** (a) Crystal structure and superconducting characteristics of $Bi_2Sr_2CaCu_2O_{8+\delta}$ monolayers and bulk forms. Monolayer refers to a half unit cell in the out-of-plane direction that contains two $CuO_2$ planes, separated by vdW gaps in bulk form (left inset). A high-quality intrinsic monolayer has Tc similar to that of the bulk form (88 K). Right inset: AFM image of monolayer and bilayer $Bi_2Sr_2CaCu_2O_{8+\delta}$. Scale bar is 10 μm. (b) Temperature-dependent resistivity of a monolayer $Bi_2Sr_2CaCu_2O_{8+\delta}$ with different initial doping levels p (holes per $CuO_2$ plaque) *via* annealing from high (purple curves) to low (red curves). (c) 2D plot of the resistivity of the monolayer as a function of temperature and doping level. The doping level p is determined as $p = 217\Omega/R_{\square}(T = 200 K)$. White circles indicate the superconducting transition temperature Tc. Reprinted and adapted with permission, (a-c) from [139] (Springer Nature).

Superconductivity and correlated insulators in magic-angle twisted bilayer graphene (MATBG) have been also studied extensively for JJs and superconducting interference quantum devices as they offer the flexibility and simplicity for fabricating high-quality JJ without edge disorder by electrostatic gating [150-155]. The conductivity of MATBG can be changed from insulating to superconducting *via* electrostatic gating [153]. By making the narrow gate electrodes for inducing superconducting regions separated by a short, non-superconducting weak link, a planar JJ can be created [155, 156]. However, the superconductivity and correlated states in MATBG have so far only been observed at very low temperatures (below 3 K) [157]. For many practical applications of JJ, 2D materials and vdW heterostructure architectures supporting high-temperature superconductivity must be identified. Recent observations of correlated states in twisted bilayer $MoS_2$ at room temperature seem encouraging [158].

## 4. Exciton condensate and high-temperature superfluidity

According to the BCS theory for superconductivity (SC), the critical temperature, $T_c$, at which Cooper pairs maintain quantum coherence, can never be higher than 30 K [ref. 159]. The discovery of SC in copper perovskites in the 1980s with a $T_c$ above 130 K highlighted that the BCS theory does not cover all types of superconductors and all phenomena associated with superconductivity [160, 161].

An exciton is a particle-like entity that forms when an electron is bound to a positively charged "hole" through Coulomb interactions. While electrons and holes are fermions, the exciton is a boson, a particle with integer spin obeying the Bose-Einstein statistics [162]. Thus, excitons can condense into a collective quantum state known as an exciton condensate, which exhibits properties similar to Cooper pairs in superconductors [163, 164]. Because excitons have a light effective mass, it may be feasible to achieve Bose-Einstein condensation (BEC) at relatively low densities and accessible temperatures [165]. Although evidence of exciton BEC has been reported over the last few decades in epitaxially grown III-V semiconductor quantum wells, they only exist at very low temperatures and/or high magnetic fields because of the low binding energy and short lifetime of the excitons in these systems [166-170].

Room-temperature exciton condensation and superfluidity were theoretically predicted in graphene bilayers by Min *et al.* in 2008, as it is a gapless semiconductor with nearly perfect electron-hole symmetry and stiff phase order [171], as shown in Fig. 9a and 9b. The author proposed a device structure in which two layers of graphene are separated by a dielectric medium with electron- and hole-doped top and bottom layers, respectively, by modulated double gates (Fig. 9a). The maximum possible Kosterlitz-Thouless temperature of the superfluids in the system is estimated to be above 300 K upon the distance between the two graphene layers and the external bias electric field (Fig. 9b). In 2017, with the discovery of high-quality and atomically thin insulating h-BN, Li *et al*. [172] and Liu *et al.* [173] independently fabricated devices with the same structure consisting of double bilayer graphene separated by a few-layer h-BN, which is very similar to the structure proposed by Min *et al*. Indeed, they observed the exciton condensate and superfluid phases in these structures, but only in the quantum Hall regime, where excitons are formed in a bilayer electron system under high magnetic field [172]. In this regime, the total number of electrons is less than the number of states in the lowest Landau level because a high magnetic field generates as many orbitals as the number of magnetic flux quanta. Owing to the particle–hole transformation, electrons occupy orbitals such that they always face empty orbitals (i.e. holes) in the opposite layer, leading to the formation of interlayer excitons [172, 173]. Evidence of exciton condensation is manifested by quantised Hall drag and dissipationless transport in the counterflow geometry [172]. Although the exciton condensation in these double bilayer graphene structures was robust and observed at a temperature ten times higher than that in a GaAs-based electron double layer [173], the requirement of a high applied magnetic field limits their practical application.

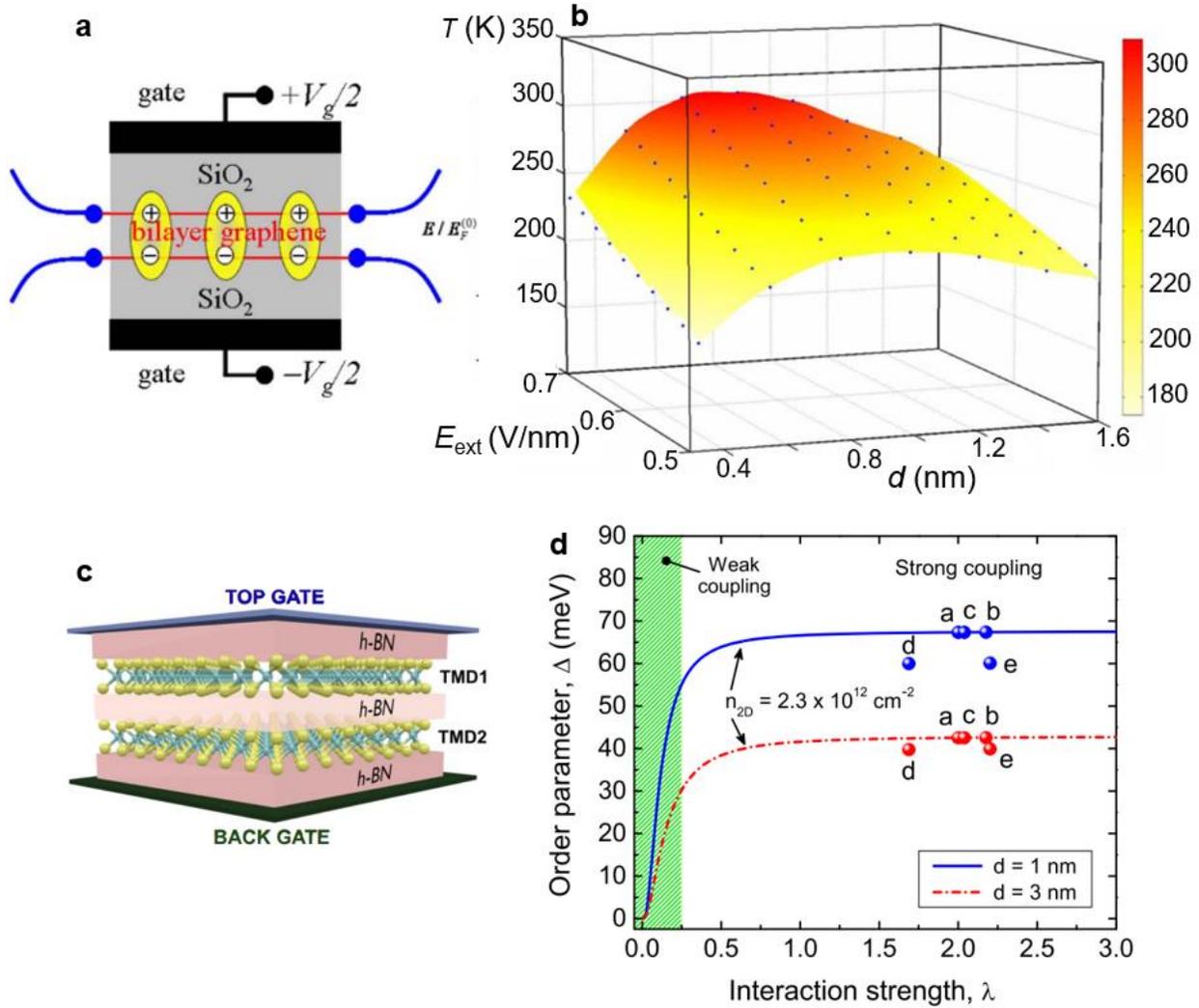

**Figure 9. Proposed vdW heterostructures for room-temperature exciton condensation and superfluidity.** (a) Two graphene layers are separated by a thin dielectric spacer while the electric field between the layers are controlled by the top and bottom gates. When the gate voltages are equal but have opposite signs ($+V_g/2$ and $-V_g/2$), the particle-hole symmetry of the Dirac equation ensures perfect nesting between the electron Fermi spheres in the n-type layer and its hole counterparts in the opposite layer, driving the Cooper instability for superfluidity. (b) Calculated superfluid phase diagram showing the dependence of the critical temperature Tc as function of the distance, $d$, and external bias field, $E_{ext}$, between the graphene layers. (c) Two monolayers of transition metal dichalcogenides (TMD) are separated by a thin insulating film of hBN. Using the double gate structure, the top and bottom layers are tuned to equal electron and hole density, respectively. (d) Calculated order parameter showing that room-temperature exciton condensation ($\Delta > 25$ meV) can be achieved in the strong coupling regime ($\lambda > 0.2$, where $\lambda$ is the ratio of the interaction energy to the band energy). $d$ (1 and 3 nm) is the distance between the two TMD layers; the spheres (red and blue colors marked a, b, c, d, e) represent the calculated results for five vdW structures including $MoTe_2/MoS_2$, $MoTe_2/MoTe_2$, $MoTe_2/MoSe_2$, $MoSe_2/WSe_2$, and $MoSe_2/MoSe_2$, respectively. Reprinted and adapted with permission, (a) and (b) from [171] (APS), (c) and (d) from [183] (APS).

Semiconducting transition metal dichalcogenides (TMDs) have excitons with long lifetimes and high binding energy; and should therefore be promising platforms for high-temperature exciton condensates in vdW heterostructures [174-183]. The exciton binding energies in monolayer TMDs are in the range of hundreds of meV [ref. 174, 177]. Long-lived interlayer excitons with lifetimes of a few nanoseconds, an order of magnitude longer than the intralayer excitons in the monolayers, have been observed in MoSe$_2$/WSe$_2$ vdW heterostructures without spacer layers [186]. Therefore, room-temperature exciton condensates have been predicted to exist in vdW heterostructures with semiconducting TMD monolayers separated by an insulating hBN spacer layer [183], as shown in Fig. 9c and 9d. The calculation of the order parameter ($\Delta$) for several vdW hetero- and homostructures, including MoTe$_2$/MoS$_2$, MoTe$_2$/MoTe$_2$, MoTe$_2$/MoSe$_2$, MoSe$_2$/WSe$_2$, and MoSe$_2$/MoSe$_2$, shows that room-temperature exciton condensation ($\Delta > 25$ meV) can be achieved in the strong coupling regime ($\lambda > 0.2$, here $\lambda$ is the ratio of the interaction energy to the band energy) when the two TMD monolayers are separated by few-layer hBN (1-3 nm) (Fig. 9d) with an exciton density $n_{2D} = 2.3 \times 10^{12}$ cm$^{-2}$. In principle all these parameters are achievable in experiments.

In fact, the strong interlayer coupling in TMD vdW heterostructures has been observed experimentally. For example, the interlayer exciton emission, where electrons in MoS$_2$ monolayer recombine with holes in WSe$_2$ layers, is still observed even with the insertion of three layers of h-BN between the two monolayers [177], as shown in Fig. 10a and 10b. Interestingly, by fabricating MoSe$_2$/hBN/WSe$_2$ heterostructure device where the electron and hole densities in the MoSe$_2$ and WSe$_2$ monolayers are controlled *via* gate voltages and interlayer bias, as shown in Fig. 10c, Wang *et al.* indeed observed evidence of high-temperature exciton condensation surviving up to 100 K [184]. While luminescence of interlayer excitons is enhanced with a narrow peak at equal electron and hole densities, the interlayer tunnelling current depends only on the exciton density, manifesting correlated electron-hole pair tunnelling (Fig. 10d-f). The critical temperatures ($T_c$) of the observed phenomenon above 100 K are consistent with the predicted exciton condensation in TMD vdW heterostructures, suggesting that it would be possible to improve Tc to approach room temperature by further optimising the device structures [183, 184].

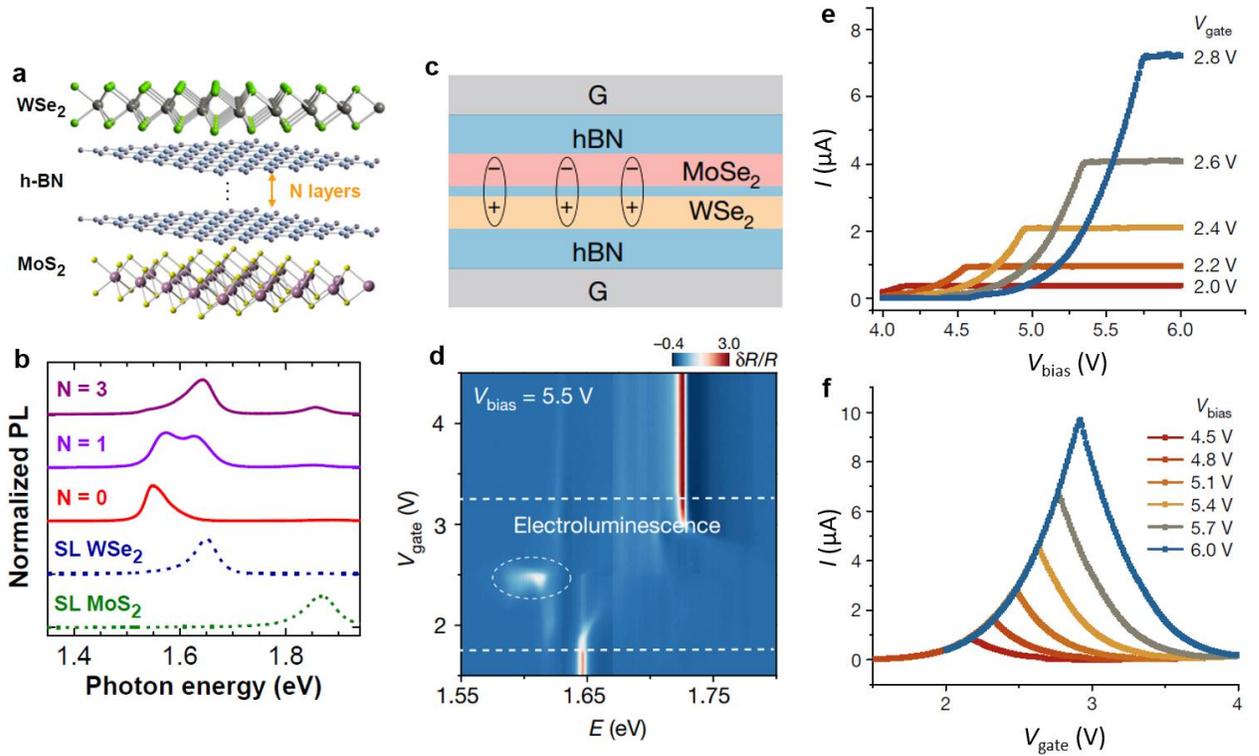

**Figure 10. Evidence of high-temperature exciton condensation in TMD-based heterostructures.** (a, b) Crystal structure and photoluminescence of $MoS_2$/$WSe_2$ with the insertion of a thin hBN insulator at the interface. The lower-energy emission peak (in red curve) from interlayer excitons formed by electrons in the $MoS_2$ monolayer and holes in the $WSe_2$ monolayer. Interlayer exciton emission is observed even with the insertion of three layers of hBN indicating strong interlayer coupling. (e-f) Device structure, electroluminescence, and electrical characteristics showing correlated electron-hole pair tunnelling, a signature of exciton condensation, in the $MoSe_2$/hBN/$WSe_2$ heterostructure. $MoSe_2$ and $WSe_2$ monolayers are separated by three-layer hBN. Few-layer graphene top and bottom gates (G) are used to tune the hole and electron densities in the $WSe_2$ and $MoSe_2$ layers, respectively. Electroluminescence of interlayer excitons is observed from the tunnel junction when $V_{bias}$ = 5.5 V is applied between the $WSe_2$ and $MoSe_2$ layers, and equal gate voltages ($V_{gate}$) are swept (circled bright spot in (d)). The tunnelling current depends only on the electron-hole pair density, but not on the individual electron and hole densities, and has a maximum value at the charge balance (n = p), manifesting the correlated tunnelling behaviour (shown in (e) and (f)). At a fixed $V_{gate}$, the current increases with increasing $V_{bias}$ and reaches a constant value at the charge balance because the density of (n + p) unchanges. Similarly, the gate-dependent tunnelling current shows a cusp-shaped peak centred at the charge balance. The phenomena are consistent with the predicted exciton condensation, and persist up to 100 K. Reprinted and adapted with permission, (a) and (b) from [177] (NAS), (c-f) from [184] (Springer Nature).

As previously mentioned, vdW heterostructures fabricated by manually stacking are commonly plagued by defects at the interfaces and angular misalignments between the layers [81-86, 102]. These imperfections weaken the interlayer coupling and thus alter the exciton lifetime and binding energy, similar to the case of interlayer tunnelling, as discussed in Section 2. Numerous studies have shown strong dependences of interlayer coupling and exciton lifetime on twist angle in TMD vdW heterostructures [185-190]. Moreover, electrons and holes located in different types of TMD layers may also induce a momentum mismatch between them leading to reduced interlayer coupling [191]. Therefore, using a natural homojunction formed within the same vdW material should, in principle, enhance the interlayer electron-hole pairing, as it provides a very clean interface structure with precisely aligned layers. We proposed a simple device architecture to search for high-temperature exciton condensates in naturally formed homojunctions within monolithic multilayer vdW crystals [192], as shown in Fig. 11. In unintentionally doped vdW layered materials, the conducting channels are spatially confined within their surface layers, while the inner layers remain intrinsic [193-197]. This allows us to build a double quantum well structure vertically within the crystal without the insertion of insulating layers [198, 199]. The spontaneously forming insulating region can be as narrow as a few nanometers in multilayer $WSe_2$, but can be tuned by the gate voltage and/or temperature [199]. By introducing holes and electrons in the top and bottom layers simultaneously by application of gate voltages, zero-bias peaks were observed in the differential conductance curves of the device at room temperature as a signature of electron-hole pairing when their densities balance (Fig. 11d). The magnitude of the conductance peak is gradually reduced under the application of a magnetic field (Fig. 11e) as it weakens the phase coherence [199-201]. Our structure closely resembles the electrical reservoir for excitons, proposed by Xie and MacDonald [202], although the inserted h-BN insulating layers are replaced with the intrinsic $WSe_2$ bulk acting as a separator. Correlated electron-hole pair tunnelling occurs because the high work function of Au (5 eV) induces a tunneling barrier for electrons injected from the source to the n-doped $WSe_2$ bottom layers, whereas holes tunnel from the drain to the p-doped $WSe_2$ top layers through the intrinsic central region.

Correlated electron-hole pair tunnelling has been previously observed as strongly enhanced coherence interlayer tunnelling between stacked graphene layers through h-BN barriers at temperatures below 10 K [203, 204]. Our observations at 300 K in multilayer crystals of $WSe_2$ open a new opportunity for the realisation of room-temperature superfluidity in vdW materials using a simple, clean, and scalable approach as large-scale multilayer 2D materials can be grown by several methods, such as CVD, PLD, or MOCVD [205-207].

Exploring novel device configurations and exotic quantum material systems is also promising for achieving high-temperature exciton condensation. For example, embedding vdW heterostructures in a cavity to induce exciton polaritons may significantly enhance $T_c$ because exciton polaritons have a much lighter effective mass [208-210]. High-temperature vdW topological insulators could also provide a suitable platform because they have well-defined surface states and insulating bulk [211-215]. Hybrid structures of 2D materials with organic or perovskite semiconductors can offer strong interlayer coupling [216-219]. Finally, moiré superlattices of vdW materials possess unique characteristics for interlayer excitons owing to the strong localisation and correlation of electrons and holes [220-225].

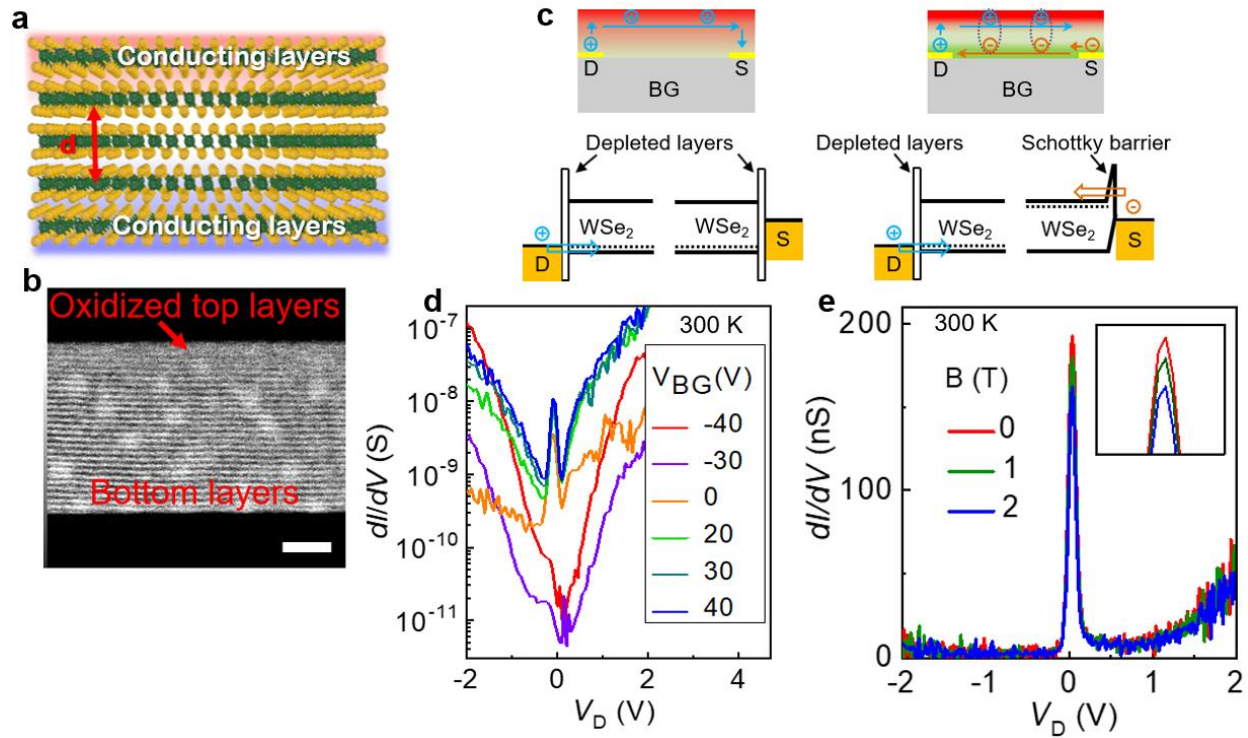

**Figure 11. Room-temperature correlated electron–hole pair tunnelling in multilayer vdW materials.** (a, b) Atomic model and cross-sectional TEM image of multilayer $WSe_2$. The top and bottom conducting layers are separated by central intrinsic insulating layers of width d. Topmost layers are unintentionally p-doped because of oxidization, whereas bottom layers could be modulated by an electrostatic back-gate. (c-e) Device configuration, band diagrams, and electrical characteristics of multilayer $WSe_2$ device with Au bottom contacts. Under an applied bias, electrons and holes tunnel into the conducting top and bottom through barriers at the $Au/WSe_2$ interfaces. Zero-bias peaks are observed in the differential conductance curves of the device only when both electrons and holes exist (positive $V_{BG}$), manifesting a signature of electron-hole pairing when their densities balance (Fig. d). The magnitude of the conductance peak is gradually reduced under the application of a magnetic field as it weakens the phase coherence (Fig. e). Reprinted and adapted with permission, (a) and (b) from [199] (Wiley), (c-e) from [192] (AIP).

## 5. Quantum spin Hall insulator and topological transistor

In normal (trivial) semiconductors or insulators, the valence and conduction bands are strictly separated by an electronic bandgap. In certain systems, the orbital structure of the crystal lattices induces an inversion of the conduction and valence electron bands, known as an inverted gap [35]. This band inversion combined with spin-orbit coupling (SOC) gives rise to the quantum spin Hall (QSH) effect, where the system has quantised spin-Hall conductance of $e^2/h$ per one helical conducting edge and vanishing charge-Hall conductance [226]. These systems are called quantum spin Hall insulators, or topological insulators [226].

Graphene was early predicted to be a 2D QSH insulator in 2005 by Kane and Mele [227]. It was later found that the gap opened by weak SOC in graphene is very small, on the order of $10^{-3}$ meV, making it unrealistic to realise experimentally [228]. The first robust QSH effect was experimentally observed in the HgTe/CdTe quantum wells in 2007 by König *et al.* [229]. Despite tremendous effort in searching for QSH insulators both theoretically and experimentally, conventional epitaxially grown materials and heterostructures and other systems such as Bi-based compounds usually have small inverted gaps of a few tens of meV, limiting the QSH states to exist at low temperatures [228-232].

Monolayer TMDs in 1T'-phase structure were proposed to host QSH states at high temperatures as it could open an inverted gap in a range of hundreds of meV as presented in Fig. 12 [233]. $MX_2$ TMDs (with M = W, Mo, and X = Te, Se, and S) exist in a variety of polytypic phase structures, including 1H, 1T, and 1T′ (Fig. 12a). While 1H-$MX_2$ monolayers are normal semiconductors with direct bandgaps, their 1T phase is typically unstable under free-standing conditions and undergoes a spontaneous lattice distortion to convert into the 1T′ structure possessing a set of quite different and unusual properties [234-237]. Using many-body perturbation theory calculations, Qian *et al.* [233] show that the structural distortion causes an intrinsic band inversion between chalcogenide-p and metal-d bands, opening an inverted gap up to 0.6 eV in 1T'-$MoS_2$ (Fig. 12b and 12c). Interestingly, this inverted gap can be modulated by an electrical field and/or strain, allowing creation of a so-called topological transistor with the possibility fast switching processes with low energy by modulation of the QSH states in the device (Fig. 12d and 12e) [233]. Robust QSH edge states are also predicted to appear at the phase boundary in various homo/heterostructures [238, 239].

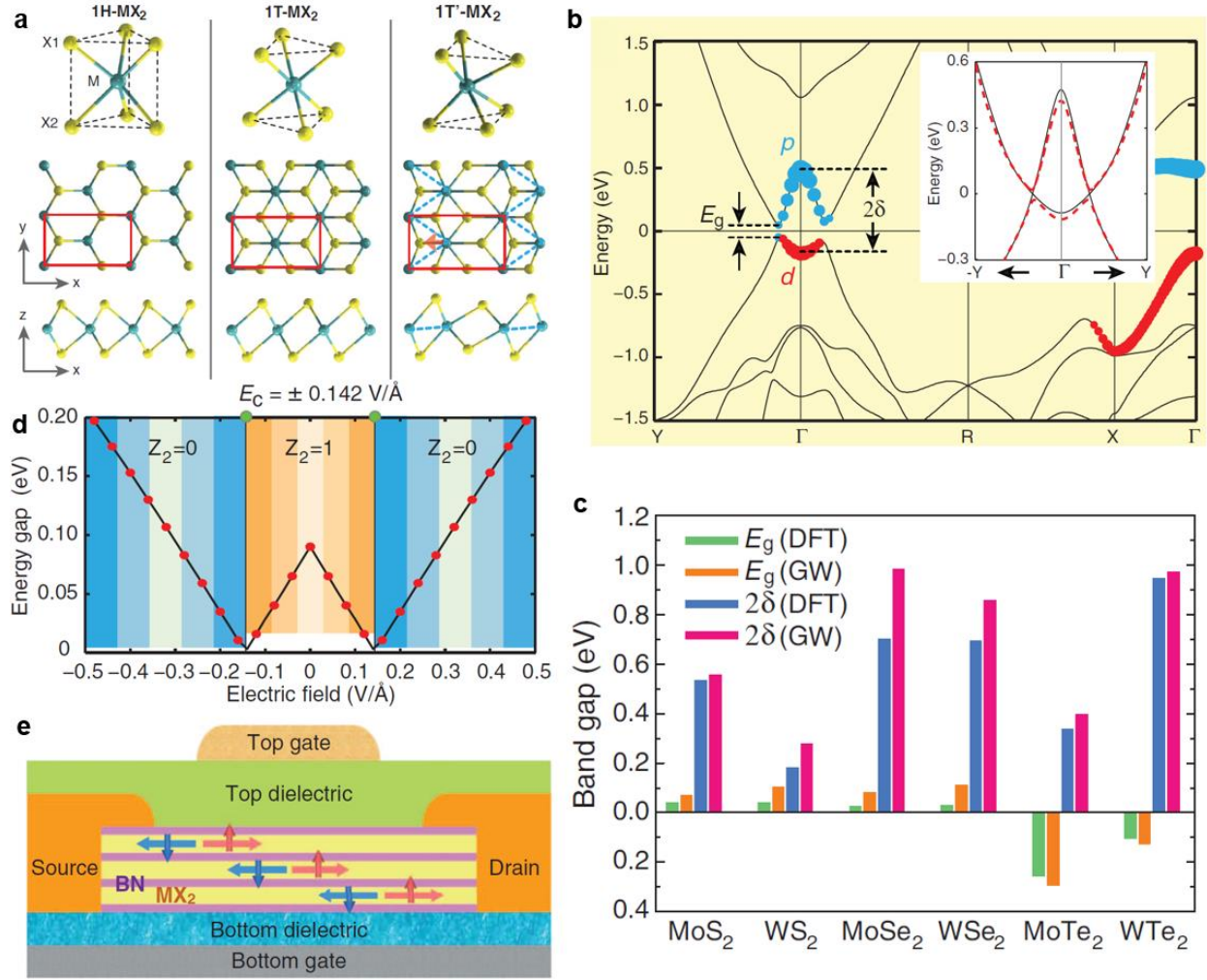

**Figure 12. Prediction of high-temperature quantum spin Hall effect in 2D materials and vdW heterostructures.** (a) Atomic structures of monolayer TMDs in different phases. 1H-MX$_2$ monolayers are normal semiconductors with direct bandgaps. The 1T phase is typically unstable under free-standing conditions and undergoes spontaneous lattice distortion to convert into the 1T' structure. (b) Calculated electronic band structures of 1T'-MoS$_2$, showing the formation of a large inverted gag (2δ) at Γ of approximately 0.6 eV. (c) Calculated fundamental gap (E$_g$) and inverted gap (2δ) of the six 1T'-TMDs. (d) Electrical field controlled topological phase diagram of 1T′-MoS$_2$. The electric field breaks the inversion symmetry and introduces strong Rashba splitting of the doubly degenerate bands near the fundamental gap E$_g$ at the Λ points. E$_g$ first decreases to zero when the electrical field increases to a critical field strength of 0.142 V/Å and then reopens. This gap-closing transition induces a topological change to a trivial phase, leading to the destruction of helical edge states. (e) The proposed vdW topological field effect transistor consists of multiple monolayers of 1T'-MX$_2$ separated by thin insulating hBN layers, preventing detrimental topological phase changes while parametrically increasing the number of edge channels. Reprinted and adapted with permission, (a-e) from [233] (AAAS).

High-temperature QSH states were observed in 1T'-MX$_2$ monolayers and at the phase boundary of 1T'-2H phase lateral junctions, as presented in Fig. 13. By carrying out the processing of the highly air-sensitive 1T'-phase TMD monolayers in an inert environment, several research groups have reported the QSH effect in 1T'-WTe$_2$ [240-243]. Wu *el al.* observed all characteristics of the QSH transport through a 2D time-reversal invariant topological insulator in the device made of 1T'-WTe$_2$ monolayer up to 100 K including: (i) the edge conductance, quantized at $e^2/h$ per edge, by helical edge modes; (ii) the saturated conductance in the short-edge limit; and (iii) suppression of conductance quantization upon application of a magnetic field owing to the loss of protection by time-reversal symmetry [240], Fig. 13a and 13b.

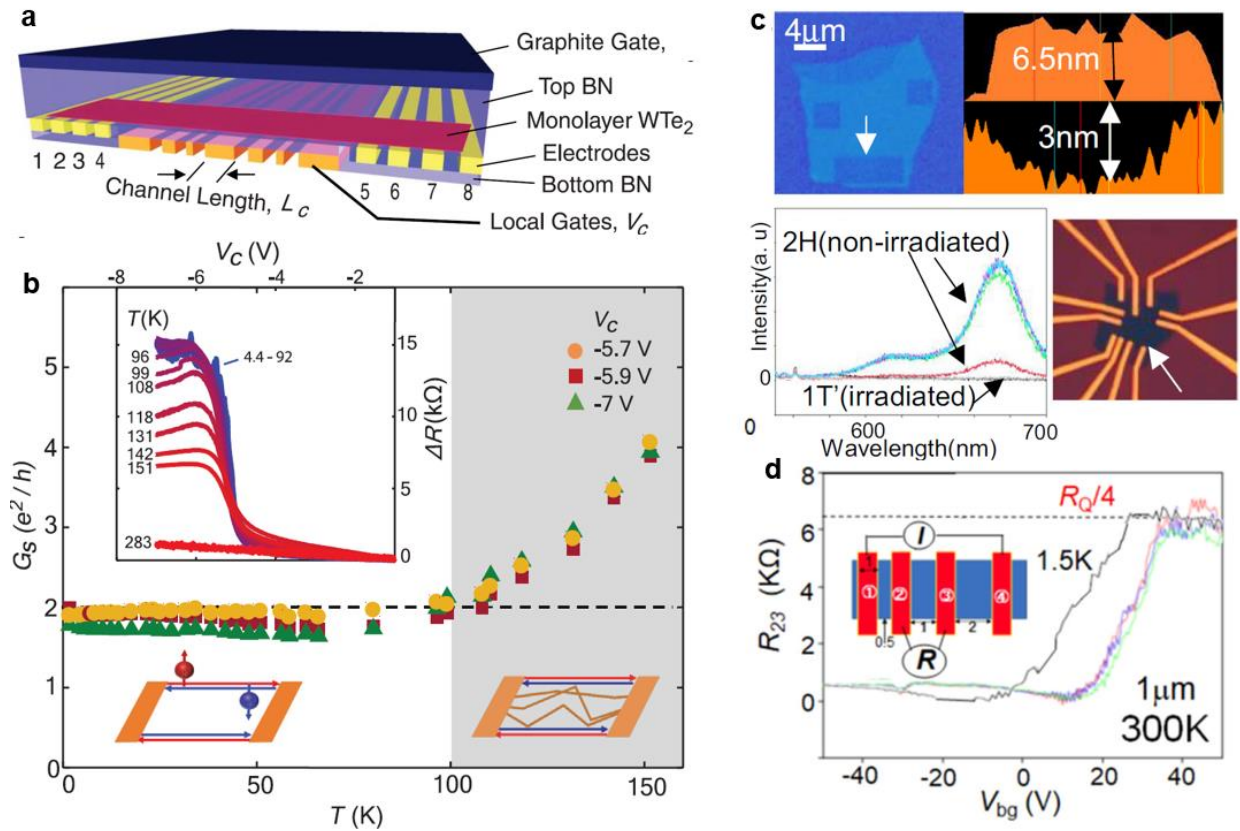

**Figure 13. Observation of the high-temperature quantum spin Hall effect in 2D materials and vdW heterostructures.** (a, b) Device structure and electrical characteristics of 1T'-WTe$_2$ monolayer device. Below 100 K, the device conductance $G_s$ remains approximately constant and close to $2e^2/h$, indicating that the conductance is dominated by quantum spin Hall effects. Above 100 K, the channel conductance increases rapidly with temperature due to the activation of bulk-conduction channels. (c, d) Evidence of room-temperature quantum spin Hall states at the phase boundary of MoS$_2$. The 1T'-phase regions are induced inside the few-layer 2H-MoS$_2$ using laser irradiation, which is confirmed by Raman spectroscopy (fig. c). The four-probe resistance of the device as a function of back-gate voltage shows plateau-like features with values of $\sim R_Q/4$, where the quantum resistance $R_Q = h/e^2$, which is similar to properties of the quantum spin Hall phase reported in HgTe quantum well systems. Reprinted and adapted with permission, (a) and (b) from [240] (AAAS), (c) and (d) from [245] (Springer Nature).

Using laser irradiation, Katsuragawa *et al.* induced a 1T'-phase region embedded in an original 2-H MoS$_2$ multilayer [244, 245], which was confirmed by Raman spectroscopy measurement (Fig. 13c). The authors then observed the saturated conductance at e$^2$/2h and e$^2$/4h, as well as the suppression of conductance quantisation under an externally applied magnetic field in the fabricated devices, as the signatures of the QSH states [245]. The phenomena survive up to 300 K, suggesting the possibility of room-temperature topological devices (Fig. 13d), although further experiments are needed to verify this. Recent advances in the growth of Mo$_x$W$_{1-x}$Te$_2$ alloys, in which their phases can be easily transformed from 2H to 1T (1T') by changing the percentage of transition metals (Mo, W), may provide a test bed for QSH states at the phase boundary [246, 247]. Introducing magnetism into a TI by adding and/or substituting atoms during the growth of 2D materials to form a magnetic TI has also recently attracted attention [248-257]. The discovery of intrinsic magnetic TI in the MnBi$_2$Te$_4$ family [252] and Kagome magnets [254-257] opens new opportunities for exploring rich quantum physics in TI systems.

The concept of electrically controlled topological phases for topological FETs has been relisied in several structures such as Bi-based TIs [258, 259], SnTe thin films [260], and few-layer Ge [261]. The search for new TIs with high Tc and innovative device configurations may enable us to achieve high-performance topological FETs operating even at room temperature [262].

Spin FET, which is usually referred as the Datta-Das spin-FET, is a conceptual spintronic device introduced by Supriyo Datta and Biswajit Das in 1990 [263]. In this device, the current flow is modulated by the electron spin rather than the charge alone. This approach can be achieved because of the Rashba spin-orbit interaction effect, a quantum phenomenon that allows electron spin to be manipulated by an electric field [264]. By tuning the gate voltage, the electron spins can be rotated by any angle between 0º and 180º as they travel through the channel. The spin current can be turned ON or OFF when the spin orientation of the electrons is aligned or misaligned with that of the drain electrode. Although spin-FETs are expected to be faster and more energy-efficient than traditional transistors, the search for semiconductors with strong spin-orbit coupling and sufficiently long spin coherence lengths to prevent the loss of spin information during transport is a great challenge [265-270]. VdW heterostructures provide new opportunities to realise high-performance spin-FETs by combining 2D materials with different functionalities simultaneously. For example, electrically controlled spin currents have been reported at room temperature in various TI/graphene and TMD/graphene heterostructures [271-275]. In these devices, the combination of the superior spin transport properties of graphene with the strong spin–orbit coupling TI or TMDs allows switching of the spin current in the graphene channel between ON and OFF states by tuning the spin absorption into TI or TMD with a gate voltage, as shown in Fig. 14a and 14b [271]. An alternative spin-tunnel FET can also be built using 2D magnetic materials, as shown in Fig. 14 c and 14d [276]. In this spin-TFET, the electrical field switches the magnetisation configuration of the 2D magnet, leading to the high- and low- tunnelling resistance states of the device [276-280].

The strong dependence of charge-to-spin conversion and spin texture generation on twist angles in graphene-based heterostructures has been theoretically predicted and experimentally observed recently [281, 282]. For example, Yang *et al.* found that changing the twist angle between graphene and $WSe_2$ layers from 7º to 36.7º reverses the helicity of the spin texture owing to the opposite unconventional Rashba-Edelstein effect [282]. This highlights again the critical role of the twist angle in the interaction of charges and spins at the vdW heterointerfaces.

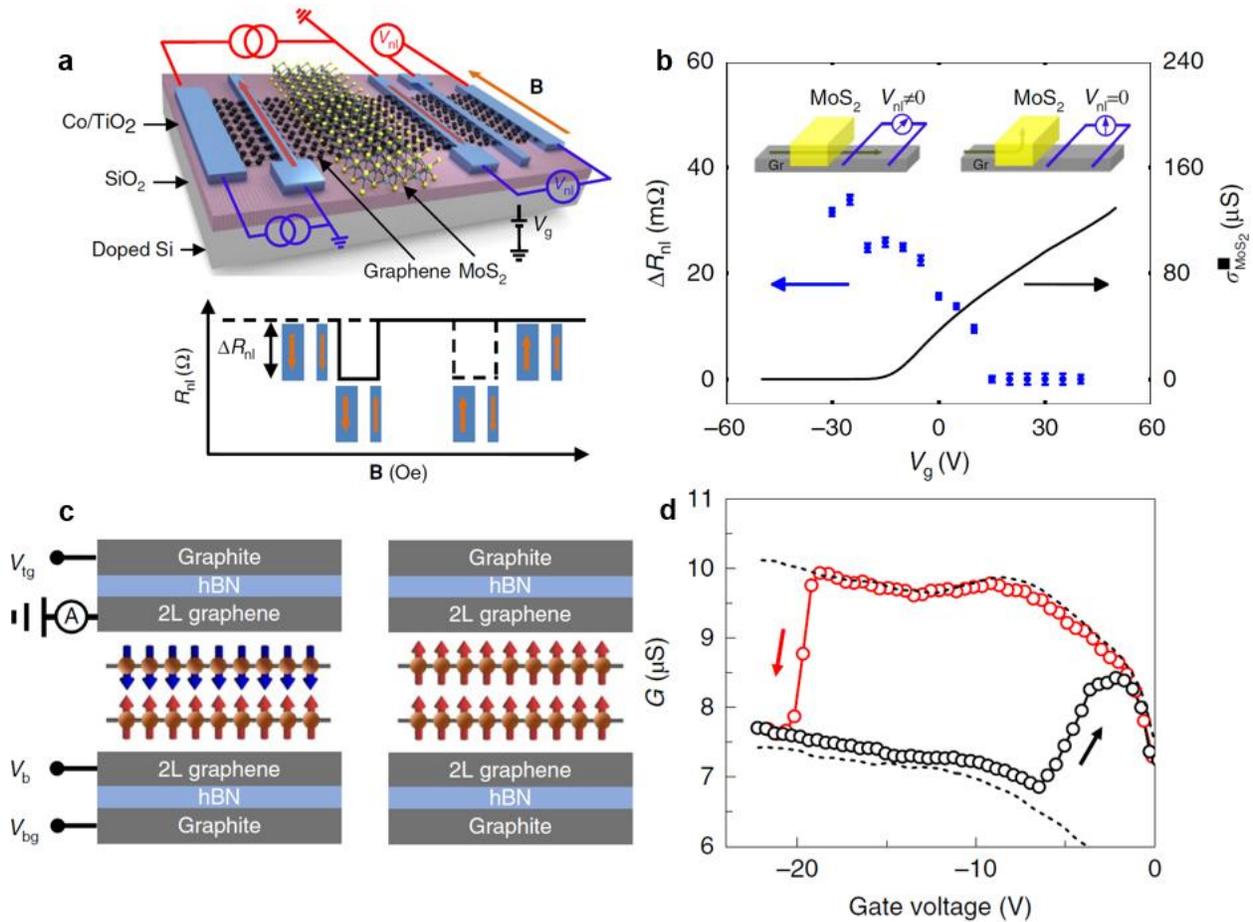

**Figure 14. vdW spin transistors**. (a) Schematic of a spin transistor using graphene/MoS2 vdW heterostructures. The detection of pure spin currents is carried out with a non-local measurement, where a DC current is injected into graphene from a ferromagnetic Co electrode across a $TiO_2$ barrier, and a non-local voltage ($V_{nl}$) is measured by a second Co electrode while sweeping the magnetic field. The spin signal is proportional to the amount of spin current that reaches the detector, which is characterized by $\Delta R_{nl}$. (b) Gate-modulated spin signal in graphene/MoS$_2$ heterostructure device (blue circles). The solid black line represents the sheet conductivity of the MoS$_2$. The insets show the spin current path (green arrow) in the OFF state (left inset) and ON conducting state (right inset) of MoS$_2$. (c) Device schematic of a spin-tunnel FET fabricated using bilayer CrI3 sandwiched between two graphene contacting layers. The operational principle of this spin-tunnel FET is based on a gate-controlled spin-flip transition and spin filtering in the tunnel junction. In the initial states (left panel), bilayer CrI$_3$ is an antiferromagnet consisting of two monolayers with antiparallel magnetic moments, resulting in high tunnelling resistance of the junction. Applying an electrical field through the top and back gates switches the bilayer CrI$_3$ to a ferromagnet through a spin-flip transition, leading to a low resistance of the junction (right panel). (d) Gate-controlled tunnel conductance of a TFET with a bilayer CrI$_3$ tunnel barrier under a constant magnetic bias of 0.76 T. Black and red symbols correspond to measurements while sweeping the gate voltage forward and backward, respectively. Reprinted and adapted with permission, (a) and (b) from [271] (Springer Nature), (c) and (d) from [276] (Springer Nature).

## 6. Topological superconductivity and Majorana modes

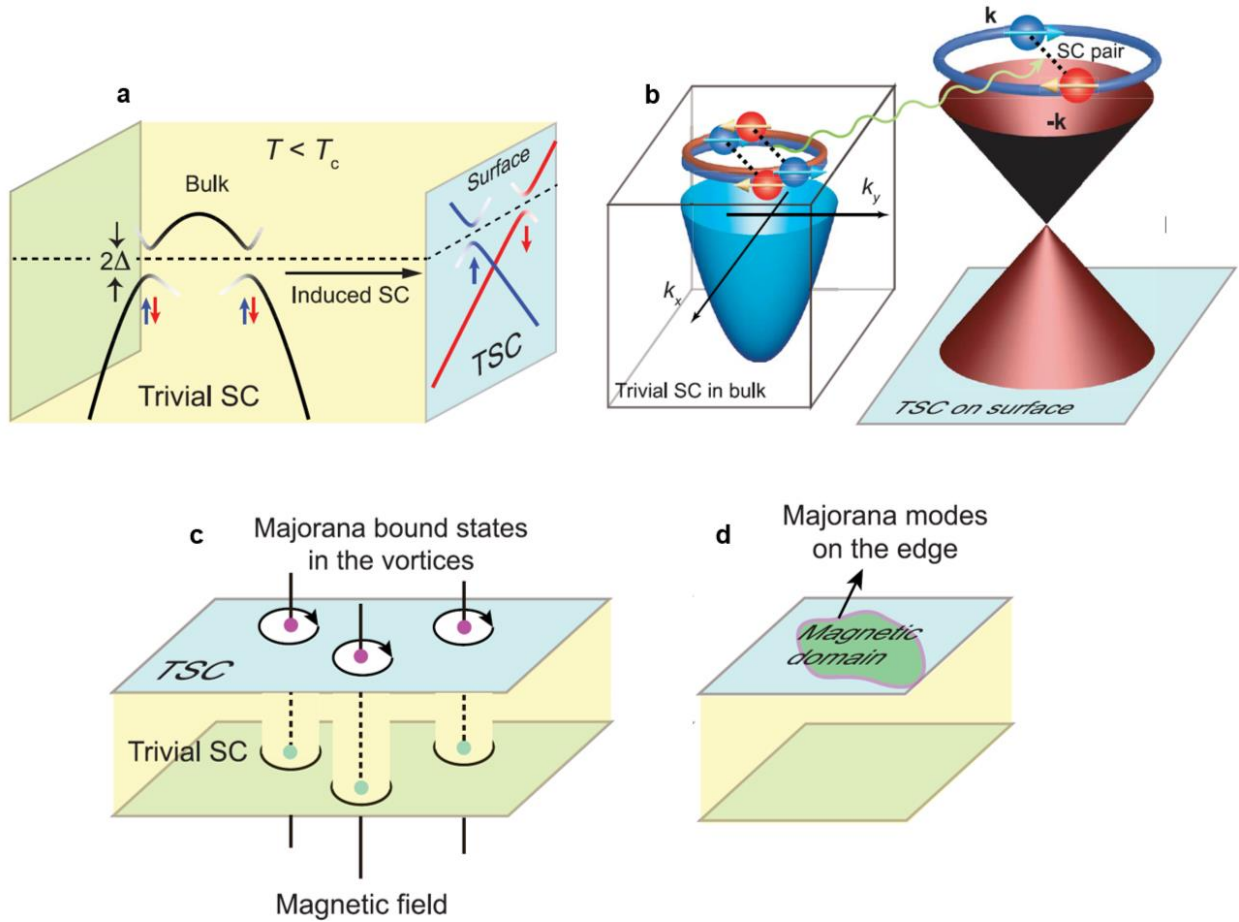

**Figure 15. Topological superconductivity and Majorana modes on the surface.** (a) Schematic visualisation of topological superconductivity in an iron-based superconductor (FeTe$_{1-x}$Se$_x$). The superconducting gaps in the bulk states are topologically trivial due to their spin degeneracy. However, the superconducting gaps of the surface states, which are induced by bulk-to-surface proximity, are topologically superconducting (TSC) because of spin polarisation. Blue and red arrows denote the spin directions. The side surface is shown for convenient visualisation. (b) Band structures with topological superconductivity on surface of FeTe$_{0.55}$Se$_{0.45}$. The electrons in the bulk are not spin-polarised, and s-wave superconducting pairing is topologically trivial. The electrons on the surface are induced to form superconducting pairs by the bulk superconductivity. The superconductivity of spin-helical surface states is topologically nontrivial. (c, d) Majorana mode formation on surface of topological superconductor. An applied magnetic field creates vortices, which behave as boundaries for topological superconductivity on the surface, and Majorana bound states are predicted to appear in the vortices. In the case of a magnetic domain existing on the surface that destroys the superconductivity within that domain, itinerant Majorana modes appear along the boundary of the domain. Reprinted and adapted with permission, (a-d) from [289] (AAAS).

In a topological superconducting structure, Cooper pairs can tunnel from the superconductor to the topological insulator surface owing to the superconducting proximity effect, thereby inducing a superconducting energy gap in the surface states [37]. These surface states are different from those of a single ordinary superconductor because they are not spin-degenerate and contain only half the degrees of freedom of a normal metal [37, 283]. Under certain conditions, the vortices in the superconductor may trap so-called zero-mode, spin-1/2 excitons of very low (zero) energy on the surface, forming a so-called Majorana zero-mode (MZMs) [283]. MZMs have been realised at low temperatures in epitaxially grown nanowires on 3D superconducting substrates and have been engineered for topological qubits [283-288]. Figure 15 represents the visualisation of the band structures of a topological superconductor candidate and the formation of Majorana bound states on its surface [289].

The combination of a 2D topological insulator with a 2D superconductor to form vdW heterostructures appears to be a straightforward route towards topological superconductivity. The evidence of topological superconductivity has been reported in different vdW heterostructures [290-293] such as $Bi_2Te_3$/$NbSe_2$, $WTe_2$/$NbSe_2$, $CrBr_3$/$NbSe_2$, and $MnBi_2Te_4$/$NbSe_2$. Some of these observations are represented in Fig. 16. By manually stacking 1T'-$WTe_2$ on the surface of superconducting $NbSe_2$, Lüpke *et al.* observed a proximity-induced superconducting gap in the edge region of the $WTe_2$ monolayer, as in Fig. 16a and 16b [291]. However, this gap is typically small (below 1 meV), and no zero-bias conductance peak was observed, which could be due to degradation of the surfaces and interfaces of the 2D layers during the sample processing [291]. Kezilebieke *et al.* grew a monolayer ferromagnet, $CrBr_3$, directly on $NbSe_2$ using molecular-beam epitaxy in an ultra-high vacuum to obtain a vdW heterostructure with clean edges and interfaces, Fig. 16a [292]. In this heterostructure, topological superconductivity is attributed to a combination of out-of-plane ferromagnetism, superconductivity, and Rashba-type spin–orbit coupling between the ferromagnet and superconductor layers [292]. The scanning tunnelling microscopy spectra of the heterostructure show that a strong zero-bias conductance peak appears at the edge region of the monolayer $CrBr_3$ (Fig. 16d and 16e). The observed conductance peak is consistent with the expected Majorana modes along the edge of the magnetic island in the established 2D topological superconductivity [292, 294]. These results confirm the critical role of the interface quality of the vdW heterostructure in achieving strong interfacial coupling.

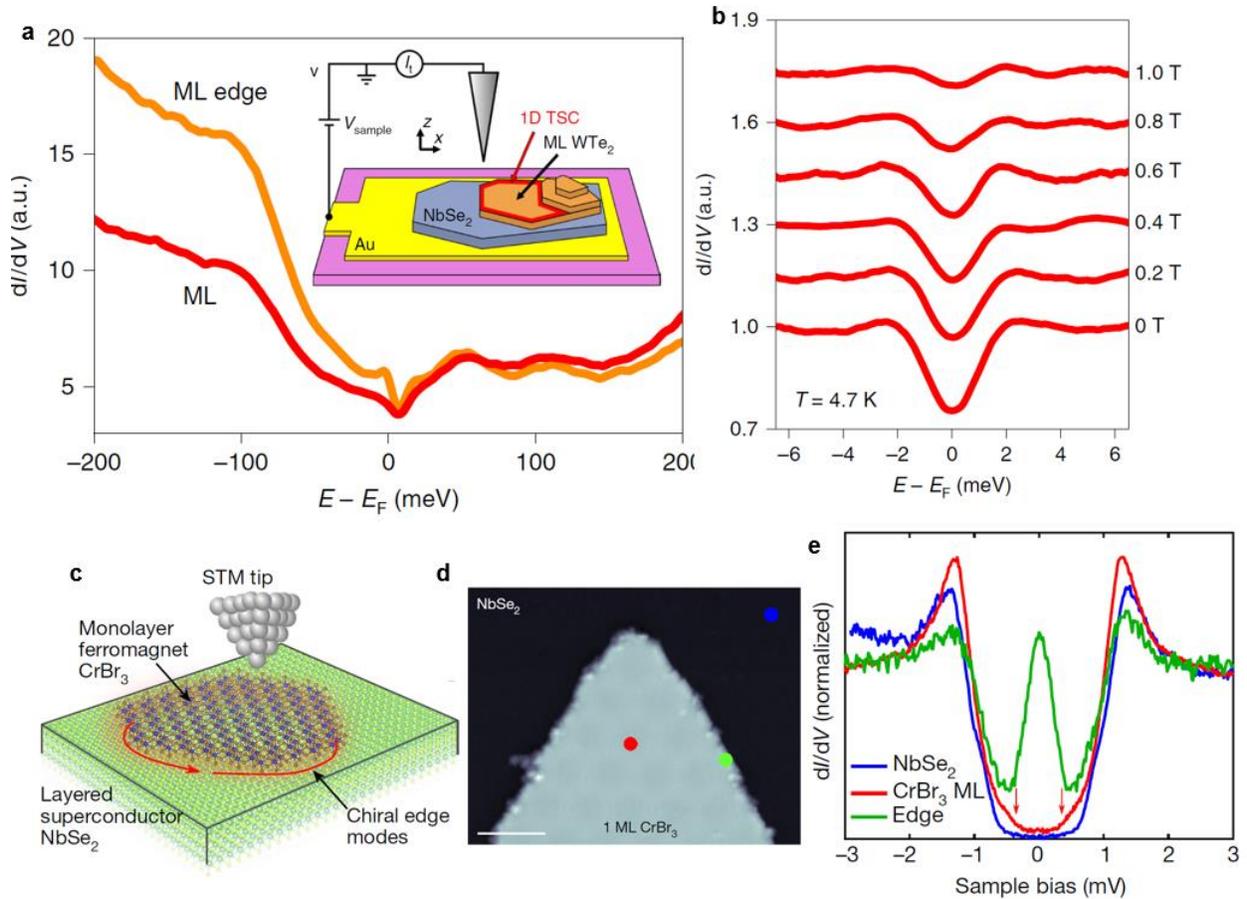

**Figure 16. Observation of topological superconductivity in vdW heterostructures.** (a, b) Differential conductance spectra of monolayer WTe$_2$ stacked on NbSe$_2$ showing the simultaneous presence of the quantum spin Hall edge state and superconducting gap on WTe$_2$. (c-e) Experimental setup, STM images, and differential conductance spectra for the realisation of topological superconductivity in the CrBr$_3$/NbSe$_2$ heterostructure. A zero-bias conductance peak appearing inside the superconductor gap is observed along the edge region of the CrBr$_3$ island. This zero-bias conductance peak is consistent with the Majorana modes calculated for 2D topological superconductors. Reprinted and adapted with permission, (a, b) from [291] (Springer Nature), (c-e) from [292] (Springer Nature).

Moiré superlattices have recently emerged as attractive systems for exploring QSH states and topological superconductors [295-299]. By fabricating twisted MoTe$_2$ homobilayers at small twist angles of 2-4 degrees, several research groups have recently reported the observation of the fractional quantum Hall effect in the absence of a magnetic field [300-304]. The observation of the fractional anomalous Hall effect in these twisted bilayers at filling factors $\nu = -2/3$ is an important and interesting achievement as it may open the opportunity to realise anyons at zero magnetic field, a key factor for stable quantum qubits [288, 300].

Motivated by the novel quantum phenomena observed in twisted vdW heterostructures, numerous device structures have been theoretically proposed to realise robust Majorana zero modes (MZMs) in these systems [148, 305-307]. We highlight some of them as in Fig. 17. Mercado *et al.* predicted that a structure consisting of a twisted bilayer of a high-$T_c$ superconductor, such as $Bi_2Sr_2CaCu_2O_{8+\delta}$ (BSCCO), on the surface of a 3D topological insulator hosts an unpaired MZM in the core of an Abrikosov vortex (Fig. 17 a) [306]. High $T_c$ of BSCCO induces potentially larger excitation gaps and higher critical temperatures of topological insulator states. In addition, the extremely short coherence length $\xi$ of this superconductor enhances MZM protection from decoherence effects [306]. Gate engineering to induce a nanowire structure on twisted bilayer graphene stacked on the surface of a TMD monolayer may also give rise to MZMs in the device (Fig. 17b) [307]. This mechanism is similar to the proximity effects observed in epitaxially grown nanowires on conventional superconductors [284]. However, MZM may be formed even at zero magnetic field, depending on the precise symmetry-breaking order [307]. The observation of the quantum spin Hall liquid characteristic in $RuCl_3$ makes it a promising material for building vdW heterostructures for the realisation of MZM [308-310]. Halász proposed a device structure for generating and detecting anyons as presented in Fig. 17c [311]. The structure consists of a single layer of Kitaev spin liquid, like $RuCl_3$, on the surface of a monolayer semiconductor, such as $MoS_2$ or $WSe_2$, and is engineered in a ring-disk-shaped device. Generation and detection of anyons are carried out through applications of gate voltages, $V_{disk}$ and $V_{ring}$, and by measuring the electrical conductance between the two leads [311].

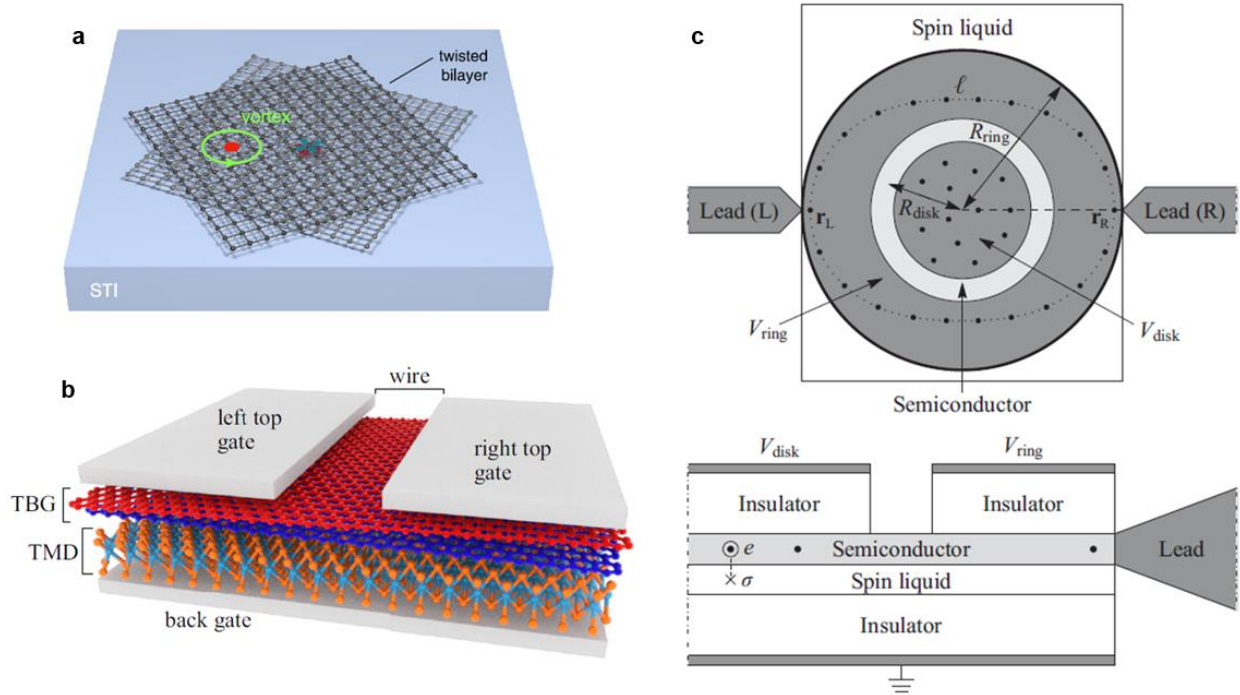

**Figure 17. Proposed vdW heterostructures for realisation of Majorana modes.** (a) Twisted bilayer high-$T_c$ superconductors stacked on the surface of a strong topological insulator generating an unpaired MZM in the core of an Abrikosov vortex. (b) Twisted bilayer graphene on TMD. The left and right top gates induce a nanowire structure on twisted graphene, and the MZM appears at the ends of this wire. The mechanism of MZM formation is similar to the proximity effects observed in epitaxially grown nanowires on conventional superconductors; however, MZM may be formed even at zero magnetic field, depending on the precise symmetry-breaking order. (c) Monolayer TMD on a spin liquid material ($RuCl_3$) with a ring-disk-shaped device. Kondo coupling stabilises an Ising anyon in $RuCl_3$ around each electron in the semiconductor. A single anyon can be generated in the disk-shaped region by gate tuning its electron number to one, while it can be interferometrically detected by measuring the electrical conductance of a ring-shaped region around it, whose electron number is allowed to fluctuate between zero and one. Reprinted and adapted with permission, (a) from [306] (APS), (b) from [307] (APS), (c) from [311] (APS).

## 7. Conclusion and perspective

In summary, we reviewed recent research progress in emergent quantum phenomena in vdW heterostructures, including quantum tunnelling of single and paired particles, exciton condensation, quantum spin Hall states, and topological superconductivity. Interface engineering is a crucial factor for achieving high-temperature-operating quantum devices. Strong interlayer coupling and tunnelling in the natural homojunction within vdW materials due to clean interfaces and perfect alignment allow the observation of quantum phenomena at high temperatures. Meanwhile, twisted bi- and multilayer structures possess numerous unexpected quantum effects which are still being intensely explored but could contribute to high-temperature quantum phenomena and devices.

As we discuss throughout the paper, the twist angle plays a crucial role in emerging quantum phenomena in vdW heterostructures but is challenging to handle in practice. First, controlling it precisely during device processing is very difficult because the ultralow friction between 2D layers easily induces instabilities and local moiré-noise [312]. Second, changing the twist angle even a fraction of a degree may lead to significant modification of moiré periodicity (in particular for low angles) and thereby the interlayer interaction, which again alters the correlations of charges and spins [102, 313]. Twist-angle programming [312] and quantum twisting microscopy [314] (see Fig. 18) have been recently reported to be useful for exploring high-temperature quantum phenomena in vdW heterostructures at arbitrary twist angles. These techniques allow rotation of the top layer at an accurate specific angle to the bottom layer with continuous ranges, while emerging phenomena from the heterostructure can be investigated *in-situ* by optical/electrical/magnetic measurements. To achieve high-temperature quantum devices with vdW heterostructure, one can start with selecting proper 2D components and then maintaining clean interfaces during device processing to search for optimized twist angle using the quantum twisting microscopy.

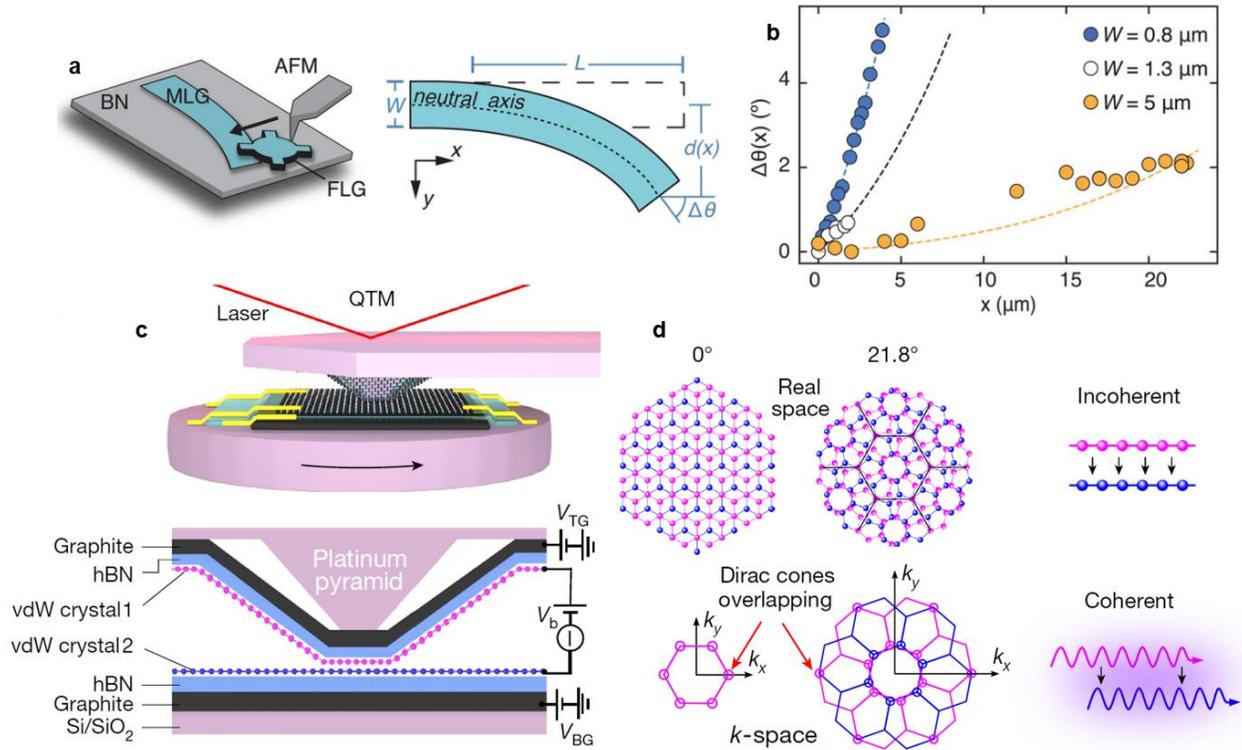

**Figure 18. Programming twist angle and quantum twisting microscopy.** (a) Schematic of bending a 2D material ribbon using a nanomanipulator and the tip of an atomic force microscope. BN: hBN substrate, MLG: monolayer graphene, FLG: few-layer graphite. The FLG manipulator is fabricated with gear-like geometry. The AFM tip slides the manipulator at the end of the MLG ribbon, inducing the bending of the ribbon. Moiré patterns with different periodicities are formed along the MLG ribbon. (b) Twist angle calculated from the measured moiré wavelength as a function of position along the ribbon with three different widths $W$ shown in (a). (c) Schematic of the quantum twisting microscopy (QTM) and experimental setup for *in-situ* electrical measurements. 2D crystals are deposited on the bottom substrate as well as on the top pyramid-shaped AFM tip located near the edge of the AFM cantilever with electrical contacts. The rotation of the AFM tip allows for continuous control of the relative angle between the tip and the sample, while the electrical properties of the heterostructure can be measured *in-situ*. (d) Real-space and momentum-space registry for commensurate angles of 0° and 21.8° between the top and bottom graphene monolayers. At a twist angle of 0º the Dirac cones of the two layers overlap at the corners of the first Brillouin zone, and at $\theta = 21.8º$ they overlap at the corners of the third Brillouin zone. Coherent tunnelling across a 2D junction can be achieved by changing the twist angle to form commensurate stackings in real space, for example, at $\theta = 21.8º$ and 38.2º for a bilayer graphene structure. Reprinted and adapted with permission, (a) and (b) from [312] (AAAS), (c) and (d) from [314] (Springer Nature).

The achievement of record low-average SS values and high on-current in TFETs at room temperature [108] and the observation of exciton condensation and quantum spin Hall above 100 K in 2D materials and vdW heterostructures [184, 240] are very encouraging and promising towards high temperature quantum devices. There are still, however, significant challenges in bringing these exotic materials and quantum functionalities into commercial applications.

High-performance devices and quantum phenomena are usually achieved only in small sample areas, where high-quality 2D materials are produced by mechanical exfoliation. For industrial applications, scalable materials with high quality and robust device processing are necessary. Research and development of wafer-scale growth [315-318], transfer techniques [319-321], pattern engineering [322-324], and 3D integration of 2D materials [325, 326] may eventually enable 2D TFETs to compete with conventional Si-based FETs.

Due to their thinness and the need to form atomically precise junctions without ultrahigh-vacuum conditions as in the case of epitaxial growth, 2D materials and their interfaces are invariably and significantly affected by experimental conditions, resulting in difficulties in terms of reproducibility, consistency, and scalability [327-328]. Today, relatively few research teams are able to fabricate high-quality twisted bilayer structures suitable for observation of quantum phenomena, such as those described in this review [329]. Therefore, the development of clear and robust protocols for vdW heterostructure processing would help accelerate research achievements, as well as more transparent, thorough, and systematic documentation of experimental methodologies and difficulties.

Contact resistance is another severe issue which limits device performance and prevents the observation of quantum phenomena, especially in TMD devices [330-335]. Despite significant efforts have been spent on this issue, achieving low contact resistance for 2D devices is still challenging, mainly because of the strong Fermi level pinning and interface disorder [331, 332]. Recent reports on the use of semimetals or charge-transfer doping for achieving low contact resistance in monolayer TMDs are promising [336-338].

Although signatures of correlated electron-hole pair tunnelling and quantum spin Hall states have been observed at room temperature [192, 245], more theoretical and experimental structures need to be attempted to obtain room-temperature superfluidity and topological superconductors. The recent discoveries of room-temperature topological insulators [339], correlated states in twisted bilayer $MoS_2$ [158], and quantum valley Hall effect in bilayer graphene [340] may provide the possibility of searching for high-temperature superconductivity and MZMs.

**Acknowledgements**

The authors thank Cheng Xiang for kindly helping with figure processing. M.-H. D. acknowledges support from the EU Graphene Flagship Core 3 (No. 881603), Villum Experiment (No. 41016), and NNF Biomag (NNF21OC0066526). P. B. acknowledges the support from DFF Metatune (1032-00496B), DFF Tr2DEO (1026-00153A), and NNF Biomag (NNF21OC0066526).